\documentclass[twocolumn]{aastex62}
\usepackage{amsmath}
\usepackage{xcolor} 
\received{\today}
\revised{\today}
\submitjournal{ApJ}

\shorttitle{DYNAMICS OF THICK SPIRALS IN PERLAS POTENTIALS}
\shortauthors{Chaves-Velasquez et al.}
\begin{document}

%\title{DYNAMICS OF THICK, OPEN SPIRALS IN PERLAS POTENTIALS - I. MODELS BASED 
%ON
%REGULAR ORBITS\footnote{Released on January, 8th, 2018}}

\title{DYNAMICS OF THICK, OPEN SPIRALS IN PERLAS POTENTIALS\footnote{Released 
on November, 2018}}

\correspondingauthor{L. Chaves-Velasquez}
\email{leonardochaves83@gmail.com}

\author[0000-0002-0786-7307]{L. Chaves-Velasquez}
\affil{Instituto Nacional de Astrof\'isica, 
\'Optica y Electr\'onica \\
Calle Luis Enrique Erro 1, 72840\\
Santa Mar\'{\i}a
Tonantzintla, Puebla, M\'exico}
\affil{University of Nari\~no Observatory\\Universidad de Nari\~no, Sede
VIPRI, Avenida Panamericana\\ Pasto, Nari\~no, Colombia}

\author{P.A. Patsis} 
\affiliation{Research Center for Astronomy, Academy of Athens \\
Soranou Efessiou 4, GR-115 27 \\
Athens, Greece}
%\collaboration{(AAS Journals Data Scientists collaboration)}

\author{I. Puerari}
\affiliation{Instituto Nacional de Astrof\'isica, 
\'Optica y Electr\'onica \\
Calle Luis Enrique Erro 1, 72840\\
Santa Mar\'{\i}a
Tonantzintla, Puebla, M\'exico}
%\affiliation{AAS Journals Associate Editor-in-Chief}
%\nocollaboration

\author{E. Moreno}
%\altaffiliation{Creator of AASTeX v6.2}
\affiliation{Instituto de Astronom\'ia, Universidad Nacional Aut\'onoma de M\'exico\\
Apdo. Postal 70-264, Ciudad Universitaria, D.F. 04510, M\'exico}
%\collaboration{(LaTeX collaboration)}

\author{B. Pichardo}
\affiliation{Instituto de Astronom\'ia, Universidad Nacional Aut\'onoma de M\'exico\\
Apdo. Postal 70-264, Ciudad Universitaria, D.F. 04510, M\'exico}

%\author{Jeff Lewandowski}
%\affiliation{IOP Senior Publisher for the AAS Journals}
%\affiliation{IOP Publishing, Washington, DC 20005}



\begin{abstract}
The PERLAS potential has been successfully used in many studies related with the 
dynamics of the spiral arms \textit{on} the equatorial plane of normal 
(non-barred) spiral galaxies. In the present work we extend these studies by 
investigating the three-dimensional dynamics of the spiral arms in the same type 
of potential. We consider a typical open, logarithmic, spiral pattern of pitch 
angle 25$^{\circ}$ and we examine the stellar orbits that can support it as the 
ratio of the masses of the spiral over the disk component ($M_{s}/M_{d}$) 
varies. We indicate the families of ``three-dimensional'' periodic orbits that 
act as the backbone of the spiral structure and we discuss their stability in 
the models we present. We study further the quasi-periodic and non-periodic 
orbits in general that follow spiral-supporting orbits as the $M_{s}/M_{d}$ 
ratio increases. We find that a bisymmetric spiral with 25$^{\circ}$ pitch angle 
is better supported by orbits in models with $0.03\lessapprox M_{s}/M_{d} 
\lessapprox 0.07$. In these cases a strong spiral pattern is supported between 
the radial 2:1 and 4:1 resonances, while local enhancements of the imposed 
spirals are encountered in some models between 4:1 and corotation. A 
characteristic bar-like structure  is observed in all models at radii smaller 
than the radius of the 2:1 resonance.
\end{abstract}



\section{Introduction} \label{sec:intro}
The PERLAS potential has been extensively used in many works that study the
dynamical properties of the spiral arms on the equatorial plane of disk galaxies
\citep{2003ApJ...582..230P, 2004MNRAS.350L..47M, JKAS,
2008ApJ...674..237A, 2010A&A...513A..51B,
2011MNRAS.418.1423A, 2012AJ....143...73P, 2012ApJ...745L..14P,
2013ApJ...772...91P, 2014ApJ...793..110M, 2015ApJ...802..109M,
2015MNRAS.451..705M, 2015ApJ...809..170P, 2016MNRAS.463..459M, 2016ApJ...817L...3M}. 
All these works shed light to the basic  
orbital dynamics in normal (non-barred) spiral galaxy models. Despite the fact
that the regions of galactic disks where the spirals exist are thin, rendering
the two-dimensional (2D) modeling as reliable, galaxies are intrinsically
three-dimensional (3D) objects. Thus, additional dynamics introduced by orbital 
instabilities due to vertical perturbations, the role of the vertical
resonances present in the disk, as well as the regular or chaotic character of
the orbits that can be used for building 3D spiral arms, should be investigated
taking into account the existence of the third dimension.

The PERLAS potential is suitable for such a study. It is 3D by construction and 
its density remains positive everywhere in the parameter space 
\citep{2003ApJ...582..230P}. This is an important advantage of the present 
model, compared with other 3D potentials used in the past for the same purpose 
\citep{1996A&A...315..371P}, since our goal is to trace the differences in the 
dynamics of models when one or more parameters vary. In PERLAS
the 
spiral arms are introduced in a different way than in the models typically employed
in literature.
Namely, in this case, the spiral arms potential is not an ad hoc perturbation 
represented by a simple function (or the addition of several of them). Instead, 
PERLAS is based on a mass density distribution that shows the intricacies and 
complications of a more realistic large scale spiral mass distribution.

There are many issues that have to be addressed for constructing a 3D spiral
pattern. The basic property of the model is to harbour spiral supporting orbits
with the appropriate morphology. Such orbits, when projected on the equatorial
plane, should precess with respect to each other as their energy varies, in such
a way that their apocentra remain close to the imposed potential minima. This
way the density will be enhanced along the spiral arms. Typical figures,
frequently used to describe this configuration, can be found in \citet[][his
figure 3]{1973PASAu...2..174K} and in \citet[][their figure
9]{1986A&A...155...11C}. In this consideration of the spiral structure, the
orbits are in agreement with the basic principle of the classic density wave
theory \citep{1964ApJ...140..646L}. However, when the amplitude of the
perturbation increases, non-linear phenomena become important and orbital
instabilities, as well as considerable morphological deviations of the periodic
orbits from ellipses may appear. According to \cite{1988A&A...197...83C} and
\cite{1991A&A...243..373P}, large amplitudes are inevitable for modeling strong,
open, spiral patterns. In the present paper we want to investigate the
limitations that are imposed by the strength of the perturbation for building 3D
spiral arms. The goal of our study is to explore the orbital dynamics of normal,
open, thick spirals, that can be considered of Sc morphological type. Another
constrain for building 3D spiral arms is their thickness. Since we do not see
morphological features attributed to the spirals exceeding the disks of spiral
galaxies observed edge-on, one has to assume a maximum height when considering
orbits. The orbits should not exceed the thickness of the disk 
at any distance from
the center.
 
Investigating the 3D spiral structure in a case of an open, bisymmetric spiral
using PERLAS potentials is ideal for a detailed study of the effects introduced
in the system when the strength of the spiral varies. In our work we study
orbits that support a spiral pattern starting at the 2:1 resonance, inside
corotation. Thus, in the present paper we do not investigate in detail the role
of chaotic orbits that start at the unstable Lagrangian points region and extend
beyond corotation \citep[see][and all relevant papers on the subject by these
groups
that followed]{2006MNRAS.369L..56P,2006MNRAS.372..901V,2006A&A...453...39R}.
However, a possible collaboration of the two mechanisms should be further
investigated, as in some models the presence and synergy of both of them can
reproduce morphologies encountered in some grand-design spirals
\citep{2017Ap&SS.362..129P}. 

We briefly present the potential components in Section~\ref{sec:model} and the
algorithms we use to calculate the periodic orbits and their stability in
Section~\ref{sec:floats}. Then, in Section~\ref{sec:periodic} we study
successively for models with $M_{s}/M_{d} = 0.01, 0.04$ and 0.07, the 2D
and 3D periodic orbits that are the backbones of each model and we discuss their
origin and their stability. The orbital content of the models based on
quasi-periodic and non-periodic orbits, as well as response models describing
their overall morphology are presented in Section~\ref{sec:orbitals}. In
Section~\ref{sec:respo} we present PERLAS response models that summarize and
verify the orbital analysis. Finally we enumerate our
conclusions in Section~\ref{sec:concl}.

\section{Potential and parameters used}\label{sec:model}
The potential model includes an axisymmetric component that has three parts.
For the central mass distribution (bulge) we adopt the spherically symmetric 
version of the \citet{1975PASJ...27..533M} potential, namely, in usual 
Cartesian coordinates,
\begin{equation}
 \Phi_{cm}(x,y,z) = -\frac{GM_{c}}{\sqrt{x^{2}+y^{2}+z^{2}+b_{c}^2}},
\end{equation}
where $M_{c}$ is the mass of the central bulge, $b_{c}$ is a scale length parameter, and $G$
is the gravitational constant.  
%A 3D  Miyamoto \& Nagai (1975) disk

For the 3-dimensional disk we use again the \cite{1975PASJ...27..533M} model, 
this time in its general form, i.e.
\begin{equation}
 \Phi_{D}(x,y,z) = -\frac{GM_{D}}{\sqrt{x^2+y^2+(A+\sqrt{z^2+B^2})^2}},
\end{equation}

where $M_{D}$ is the mass of the disk and $A$, $B$ are scale lengths.

For the massive halo we used the potential of \cite{1991RMxAA..22..255A}, 
which at radius $r$ is given by
\begin{eqnarray}
\Phi_{H}(r)=-\left(\frac{M(r)}{r}\right)-\left(\frac{M_{H}}{1.02a_{H}}
\right)\times\nonumber \\ 
 \left[-\frac{1.02}{1+(r/a_{H})^{1.02}}+\ln(1+(r/a_{H})^{1.02})\right]^{100}_{r}
, 
\end{eqnarray}
where
\begin{equation}
 M(r) = \frac{M_{H}(r/a_{H})^{2.02}}{1+(r/a_{H})^{1.02}}.
\end{equation}
Here $M(r)$ has mass units, $M_{H}$ is the mass of the halo, and $a_{H}$ is a
scale length.  

Superposed to the axisymmetric components, a non-axisymmetric component is 
included that represents the spiral arms. For this component, we used the PERLAS 
potential \citep{2003ApJ...582..230P}. It is a bisymmetric,  
three-dimensional potential and is shaped by a density distribution formed by 
%%%%%%%%%%%%%%%%%%%%%%%%%%%%%%%%%%%%%%%%%%%%%%%%%%%%%%%%%%%%%%%%%%%%%%%%%%
individual, inhomogeneous, oblate Schmidt spheroids \citep{1956BAN....13...15S},
superposed in this study along a 
logarithmic spiral locus of constant pitch angle $i$. The spirals are 
considered to unwind clockwise.
% individual, inhomogeneous, oblate spheroids, superposed along a logarithmic 
% spiral locus. The phase of the spiral at radius $r$ is given by the form 
% \begin{equation}
%  f(r) = 
% -\frac{2}{N\tan{i_{0}}}\text{ln}\left[1+\left(\frac{r}{r_{s}}\right)^{N}\right]
% \end{equation}
% which has been suggested by \cite{rhva79} and contains information
% about the geometry and the direction of the spiral pattern. The parameter 
% $r_{s}$ 
% is the galactocentric distance where the spiral pattern begins and $N$ 
% determines the statring points of the spiral at the $r_{s}$ radius. The minus 
% sign in this equation implies that the spiral arms unwind clockwise, while 
% $i_{0}$ is the constant pitch angle.

% shape of the spiral arms at the begining, such that for $N=0$
% the spiral arms start with a $180^{o}$ degrees, and for $N\rightarrow\infty$ the																				
% spirals start with $90^{o}$ angle. The minus sign in this equation implices that
% the spiral arms roll clockwise, and $i_{0}$ is the pitch angle.

The spheroids used to model the spirals have 
constant semi-axes ratio and their density falls linearly outwards, starting
from their centers on the spiral locus. Their 
total
width and height are 2 and 1 kpc,
respectively; the separation among the centres of the spheroids is 0.5 kpc. The
spiral arms formed this way begin at the ILR and end at a distance $1.5$ times
the corotation radius. Their density along the locus falls exponentially, as  
the one of
the disk does. We consistently add the spiral arms mass subtracting it from the
disk to maintain the model invariable in mass. For details on the spiral arms
model PERLAS, see for instance
\citet[][]{2003ApJ...582..230P,2009ApJ...700L..78A, 2012ApJ...745L..14P,
2013ApJ...772...91P}.

Having in mind to model the open spirals of an Sc type galaxy, we adopted the
parameters used for that purpose by \citet{2015ApJ...809..170P}.
We briefly summarize them in Table~\ref{tab:parameters}. We give the parameters for the
axisymmetric components, as well as those for the spiral perturbation.
%\textbf{PUT HERE TABLE 1 - In the caption please say that it corresponds to the
%table in PV}
\begin{table*}%[h!]
\begin{center}
\centering \footnotesize \caption{Parameters of the galactic models.}
\label{tab:parameters}
\scalebox{0.8}{
\begin{tabular}{ccccccc}

\hline
Galaxy type    & Spiral Locus & Arms Number & Pitch Angle 
$i^{o}$ & $\mu=M_{sp}/M_{D}$ & Scale length (kpc) & $\Omega_{p}$ 
($\text{km}/\text{s}/\text{kpc}$)\\[0.2cm]
%\hline
  Sc    & logarithmic    & $2$  & $25^{o}$ & $0.01-0.10$ & 
$3.7$ & $-20$    \\

\hline\\
 
\multicolumn{7}{c}{Axisymmetric Components} \\
\cline{1-2}

\hline

%\begin{center}
$M_{D}/M_{H}$    & $M_{c}/M_{D}$ & Rotation Velocity\footnote{Maxima of the
rotation velocity.} ($\text{km}\text{s}^{-1}$) & $M_{D}(10^{10}M_{\odot})$ &
$M_{c}(10^{10}M_\odot)$ & $M_{H}(10^{11}M_\odot)$ & Disk Scale-length (kpc) 
\\[0.2cm]
%\hline
$0.1$     & $0.2$    & $170$  & $5.10$ & $1.02$ & $4.85$ & $3.7$      \\
%\end{center}
\hline
 & $b_c$ (kpc) & $A$ (kpc) & $B$ (kpc) & $a_H$ (kpc) &  \\ [0.2cm]
 &   1         &   5.32    &  0.25     &    12       &  \\
\hline 
%\multicolumn{7}{c}{Constants of the Axisymmetric Components in PERLAS units\footnote{dhsgdhgsdh}} \\
%\cline{1-7}
%Bulge($M_{B}$, $b_{1}$) & Disk ($M_{D}$, $a_{2}$, $b_{2}$) & Halo ($M_{H}$, $a_{3}$) \\[0.2cm]
%$400$, $1.0$ & $2200$, $5.3178$, $0.25$ & $2800$, $12.0$ & \\
%\hline
\end{tabular}
}
\end{center}
\end{table*}
The rotation curve and its decomposition is given in bottom panel of figure 1 in
\citet{2015ApJ...809..170P}. The two most characteristic parameters of our model 
are the pattern speed
$\Omega_p=-20$~km s$^{-1}$ kpc$^{-1}$ with which our system rotates and the
pitch angle of the logarithmic spirals, for which we have adopted the value
$i=25^{\circ}$, typical for an Sc galaxy. The amplitude of the spiral
perturbation is determined by the ratio $M_{s}/M_{d}$. This is the parameter we
have varied in our study in the range $0.01 \lesssim M_{s}/M_{d} \lesssim 0.1$. 

\begin{figure}
\epsscale{1.0}
\plotone{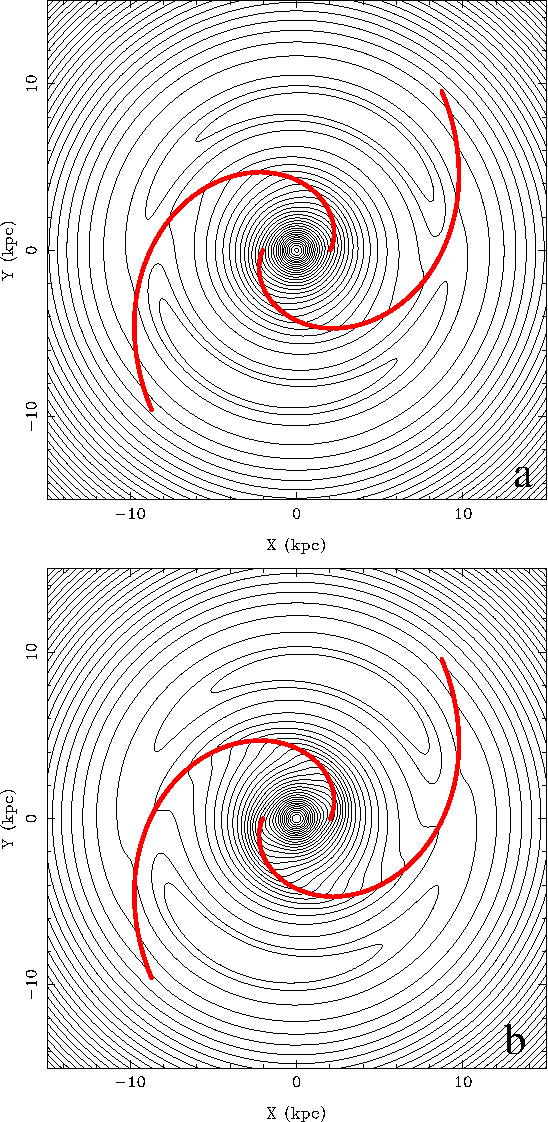}
\caption{The effective potential on the $(x,y)$ 
equatorial plane of the PERLAS model we use. (a) in the case $M_{s}/M_{d}=0.04$, 
(b) in the case $M_{s}/M_{d}=0.1$.  }
    \label{effpot}
\end{figure}

An informative diagram about the properties of the potential is 
Fig.~\ref{effpot}. In this figure we give the effective potential in the 
equatorial plane $(x,y)$ in the cases with $M_{s}/M_{d}=0.04$ (a) and  
$M_{s}/M_{d}=0.1$ (b). The first case corresponds to one of the standard models
of 
this study, which will be discussed in detail in the following sections. The 
second is the model with the most massive spirals we studied, so the properties 
of the imposed potential are displayed pronounced.

Estimating the basic radial resonances of the system from the axisymmetric part 
of the model, corotation is at $8.63$~kpc, the 2:1 resonance at $2.03$~kpc and 
the 4:1 resonance at $5.35$~kpc. (Thus, in our models the 
spiral arms end at  12.9~kpc, which is 1.5 times the corotation radius).
 However, small displacements are expected to 
occur when the perturbation 
is introduced, especially when the spiral is strong. This will be discussed for 
each model separately. In addition, in our model we have also vertical 
frequencies, in the $z$ direction, defined by 
\begin{displaymath}
 \nu^2 (R)=\left( \frac{\partial^2 \Phi_0}{\partial z^2} \right)_{z=0}, 
\end{displaymath}
where $\Phi_0$ is the axisymmetric potential,  with the corresponding vertical
resonances determined by 
\begin{displaymath}
 \frac{\nu}{\Omega - \Omega_p}=\frac{n}{m}.
\end{displaymath}

In practice the location of the radial and vertical resonances should be always
taken from the variation of the stability curves of the
spiral-supporting families of periodic orbits, since these curves have specific
properties at the resonances (see Sections ~\ref{sec:floats} 
and \ref{sec:periodic} below).

\section{Periodic orbits and stability indices} \label{sec:floats}

In order to study the orbital behavior of our galactic models, we use the
Hamiltonian formalism. As in the previous, 2D, PERLAS studies, the galaxy is
modelled as an autonomous Hamiltonian system, rotating with pattern speed
$\Omega_{p}$, in which case the Hamiltonian can be written as 
\begin{equation}
 H =
\frac{1}{2}\left(p_{x}^2+p_{y}^2+p_{z}^2\right)+\Phi(x,y,z)-\Omega_{p}(xp_{y}
-yp_{x})=E_{j}
\end{equation}
where $x,y,z$ are Cartesian coordinates in the rotating frame,
$p_{x},p_{y},p_{z}$ are the canonically conjugate momenta, $\Phi(x,y,z)$ is the
total potential (including axisymmetric and non-axisymmetric terms) and $E_{j}$
is the Jacobi's constant (in the text we may also refer to it as the
``energy'').

The equations of motion are
\begin{eqnarray}
  \dot{x} = p_{x}+\Omega_{b}y \nonumber\\
 %\end{equation}
 %\begin{equation}
  \dot{y} = p_{y}-\Omega_{b}x \nonumber\\
 %\end{equation}
 %\begin{equation}
   \dot{z} = p_{z} \nonumber\\
 %\end{equation}
 %\begin{equation}
  \dot{p}_{x} = -\frac{\partial\Phi}{\partial x} + \Omega_{b}p_{y} \\
 % \end{equation}
 % \begin{equation}
  \dot{p}_{y} = -\frac{\partial\Phi}{\partial y} - \Omega_{b}p_{x} \nonumber\\
 % \end{equation}
 % \begin{equation}
   \dot{p}_{z} = -\frac{\partial\Phi}{\partial z}\nonumber
 \end{eqnarray}

In order to calculate orbits we start in the plane $y=0$ with $p_{y} >0$. Due to 
the existence of the $E_{j}$ integral the needed initial 
conditions can
be reduced in four $(x_{0},p_{x_{0}},z_{0},p_{z_{0}}$). Our system rotates 
clockwise, thus the value of the initial $x_{0}$ coordinate will be negative. 
For finding periodic orbits we use an iterative Newton method with  an accuracy 
of at least $10^{-10}$. The equations of motion for this purpose are integrated 
using a Runge-Kutta integrator of order four. For the integration of chaotic 
orbits and in the calculations of Poincar\'{e} surfaces of section we have used 
also a Runge-Kutta Fehlberg 7-8$^{th}$ order, scheme \citep{f68}.

The computation of the stability indices of the periodic orbits is based on the
theory of variational equations. We consider small deviations from its initial
conditions, and then the orbit is integrated. Therefore the initial and final
deviation vectors are related as:
\begin{equation}
\boldsymbol{\xi} = M \boldsymbol{\xi}_{0},
\end{equation}

where $M$ is the monodromy matrix. The characteristic equation of this matrix has the form
\begin{equation}
 \lambda^{4}+\alpha \lambda^3+\beta \lambda^2+\alpha\lambda+1=0
\end{equation}
Its solutions obey the relations $\lambda_{1}\lambda_{2}=1$,
$\lambda_{3}\lambda_{4}=1$ and for each pair we can write
\begin{equation}
 \lambda_{i},1/\lambda_{i}=\frac{1}{2}[-b_{i}\pm(b_{i}^2-4)^{1/2}]
\end{equation}
 where $b_{i}=\frac{1}{2}(\alpha\pm\Delta^{1/2})$ and
$\Delta=\alpha^2-4(\beta-2)$. $b_{1}$ and $b_{2}$ are the stability indices. If
$\Delta>0$, $|b_{1}|<2$ and $|b_{2}|<2$, the four eigenvalues are on the unit
circle and the orbit is stable. If $\Delta>0$, $|b_{1}|>2$ and 
$|b_{2}|<2$ or
$|b_{1}|<2$ and $|b_{2}|>2$, two eigenvalues are on the real axis and two on the
unit circle, and the orbit is simple unstable. If $\Delta>0$, $|b_{1}|>2$ and
$|b_{2}|>2$, all four eigenvalues are on the real axis and the orbit is called
double unstable. Finally if $\Delta<0$, all four eigenvalues are complex 
numbers
but off the unit circle and then the orbit is characterized as complex 
unstable. For a detailed
description of the method the reader is referred to 
\cite{br} and \cite{1975CeMec..12..255H}.

\vspace{0.5cm}
\section{2D and 3D periodic orbits as building blocks}\label{sec:periodic}
Sketches like the one by \citet[][his figure 3]{1973PASAu...2..174K} illustrate
the assumed stellar (and to a large degree also gaseous) flows in a barred and
in a spiral case. In 2D models the drawn ellipses correspond to elliptical
periodic orbits (hereafter p.o.) that belong to the x1 family
\citep{1989A&ARv...1..261C}. In cases in which the elliptical p.o. have their
apocentra along an axis, they support a bar, while in cases they have their
apocentra close to spiral loci, they support a spiral pattern. Essentially, in
both cases, it is the same family of p.o. found in different potentials. A
bisymmetric spiral pattern can be seamlessly reinforced in such a configuration
between the radial 2:1 and 4:1 resonances. Beyond that resonance the ellipses
obtain a rectangular character that leads either to boxy bars
\citep{1980A&A....81..198C, 1997ApJ...483..731P}, or to bifurcations of the arms
and/or an overall deterrence of the evolution of open, 
bisymmetric stellar
spirals in normal spiral
potentials \citep{1986A&A...155...11C, 1991A&A...243..373P}.
	
In addition, in a 3D case, the x1 family is substituted by a tree of families of
p.o., called by \citet{2002MNRAS.333..847S} the ``x1-tree''. The x1-tree
includes, besides the planar x1 family on the equatorial plane, also the
vertical bifurcations of x1, which are introduced in the system in pairs at the
vertical resonances, starting from the vertical 2:1 resonance. In
\citet{2002MNRAS.333..847S} they have been named x1v1, x1v2 (the two families
that are associated with the vertical 2:1 resonance), x1v3, x1v4 (the two
families that are associated with the vertical 3:1 resonance), etc. This is the
usual succession of the 3D families of p.o. in rotating galactic models.
Exceptions have been found in particular cases \citep{2018A&A...612A.114P}, but
usually the families introduced as stable in the system are those that have as
last digit in their name in the Skokos et al. nomenclature an odd number (x1v1,
x1v3, etc.). Looking for 3D stable families to support the spiral arms, these
are
the obvious first candidates to be examined as possible building blocks of a
thick spiral. These families, being bifurcations of the planar x1 family at a
transition of stability from stable to simple unstable at a vertical resonance,
have by definition exactly at the bifurcation point a morphology identical to
x1. However, as soon as we depart from the bifurcating point towards larger
energies (Jacobi constants) they develop a typical morphology for each
family.
As in the case of barred models \citep{2002MNRAS.333..847S} the p.o. of the
x1-tree are expected to have at a certain energy range elliptical projections on
the equatorial plane, similar to the x1 ellipses. Then, gaining in height, they
will have projections that will not be able to be combined with the x1 p.o. and
support the same spirals. We have to underline though that the typical edge-on
morphologies associated with each family in barred potentials is because of
their orientation with respect to the major and minor axes of the bar, which
coincide with the corresponding axes of the elliptical orbits. In a spiral
potential however, since the axes of the 3D periodic orbits of a family
projected on the equatorial plane precess as their $E_j$ varies, their
projections on the $(y,z)$ and $(x,z)$ planes are not expected to be in
general the known, recognizable shapes of the corresponding side-on and end-on
views of the barred cases. 

We note that in a non-axisymmetric system the most reliable way of locating the
resonances, is by means of its periodic orbits. At the radial and vertical
resonances new families are introduced. Thus, one has to find the bifurcating
points along the characteristic curve, or along the stability diagram, of the
main family x1 and associate them with a $n:1$ resonance. The location of a
resonance specified this way, is expected to be different from a direct
estimation from the $\Omega \pm \lambda/n$ vs $r$ curve (where $\lambda$ is
either the epicyclic or the vertical frequency of the radial and vertical
resonances respectively), which is based on the axisymmetric part of the
potential. In weak spiral models this difference can be practically ignored, but
it has to be taken into account in strong spiral cases.  

In Table~\ref{tab:spiral} we give for the models we will
refer to in the following sections the forcing, i.e. the maximum tangential
over the axisymmetric force, at the resonances. 
\begin{table}[]
  \begin{center}
    \caption{Forcing at the main resonances for models M1, M4, M7 and M10.}
\label{tab:spiral}
\begin{tabular}{lcccc}
\hline
model & $M_{s}/M_{d}$ & ILR & 4:1 & corotation \\ 
    &  & (2.03~kpc) & (5.23~kpc)  &  (8.63~kpc)    \\ \hline
 M1 & 0.01 & 0.024 & 0.023 & 0.016 \\ 
 M4 & 0.04 & 0.097 & 0.094 & 0.064 \\ 
 M7 & 0.07 & 0.170 & 0.165 & 0.112 \\ 
 M10& 0.10 & 0.243 & 0.236 & 0.175 \\ \hline
\end{tabular}
\end{center}
\end{table}

\subsection{Model $M_{s}/M_{d}=0.01$}\label{m1}
\begin{figure}
\epsscale{1.15}
\plotone{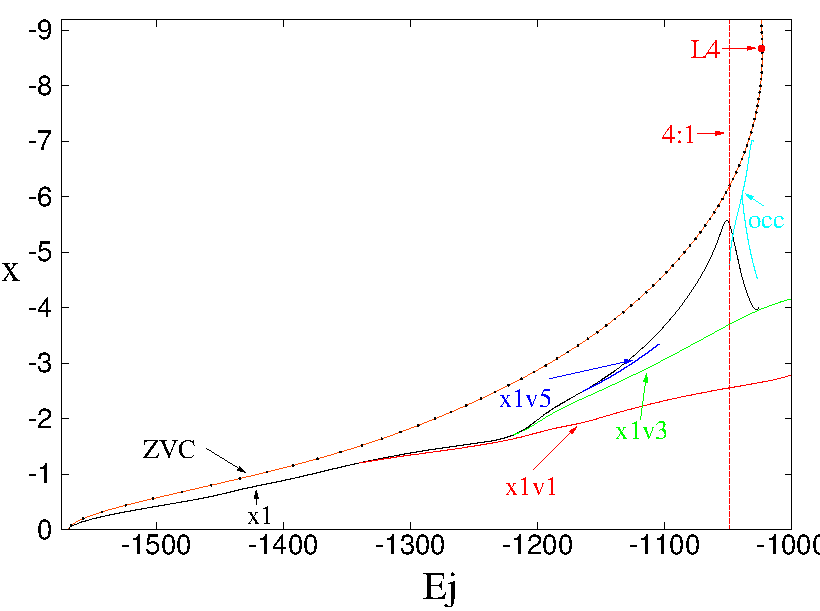}
\caption{Model M1: The $(E_j,x_0)$ characteristic curves of the x1-tree families
of the model. Indicated with arrows are given the characteristics of the
families x1 (black), x1v1 (red), x1v3 (green) and x1v5 (blue). The curves in
light blue at the upper right part of the diagram correspond to planar families
on the equatorial plane close to corotation (labeled ``occ''), which are
introduced at radial $n:1$ resonances with $n\geq 4$. The curve of zero velocity
(ZVC) is the uppermost red curve with black dots, also indicated with an arrow.
The dashed vertical line indicates the location of the 4:1 resonance and the red
dot at the local maximum of the ZVC at $E_j \approx -1022.65$, gives the L4
point.}
    \label{cc001}
\end{figure}

The less massive spiral we have examined has $M_{s}/M_{d}=0.01$. We will refer 
to it as Model ``M1''. Fig.~\ref{cc001} gives the $(E_{j}, x_0)$ 

\citep[for a definition see][]{1989A&ARv...1..261C} for the x1-tree families of 
this model. It is 
composed by the individual projections of the characteristic curves, while in 
3D 
systems the p.o. demand in general four initial conditions $(x_0, z_0, p_{x_0}, 
p_{z_0})$ 
in order to be uniquely specified and be plotted against $E_{j}$. Even for the 
characteristic curve of the  planar x1 family we need two initial conditions 
$(x_0, p_{x_0})$, because in general the x1 ellipses in a spiral model have 
$p_{x_0}\neq 0$. Nevertheless the $(E_{j}, x_0)$ projection gives information 
about the Jacobi constant in which a family is introduced in the system and its 
extent, so we will use it, together with the stability diagrams, for the 
description of the models.

In Fig.~\ref{cc001} the black curve is the characteristic of
the x1 family, the red
curve that of the 3D x1v1 family, the green curve is the x1v3 characteristic,
the blue one corresponds to the x1v5, while the curves depicted in light blue
are 2D orbits close to corotation, bifurcated at \textit{radial} $n:1$
resonances with $n\geq 4$. $E_{j}$ is given in units of $(10 km/s)^2$. The same
holds for all figures $E_{j}$ appears, throughout the paper.

\begin{figure}
\epsscale{1.15}
\plotone{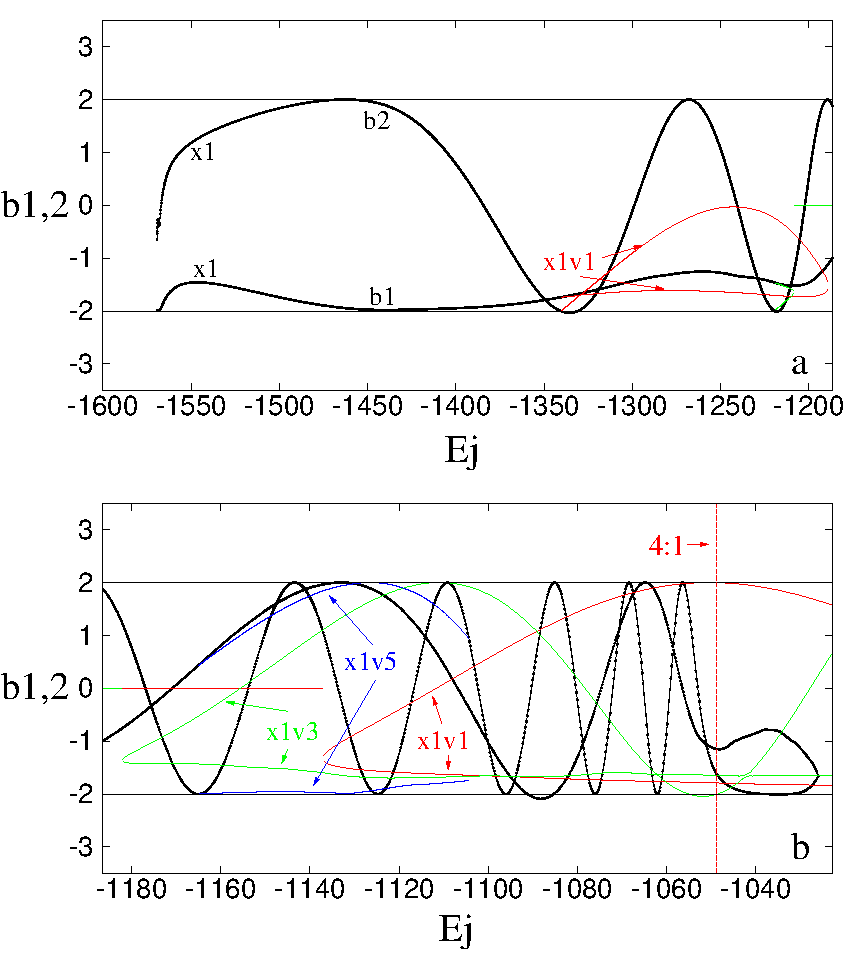}
\caption{Stability diagram for model M1. (a) in the range $-1600\le E_j\le
-1186.45$ and (b) in the range $-1186.45\le E_j\le -1022.71$. The indices are
plotted for x1 in black, x1v1 in red, x1v3 in green and x1v5 in blue. Straight
line segments at $b_{1,2}=0$ indicate the energy ranges, where a family becomes
complex unstable. The dashed vertical red line indicates the position of the 4:1
resonance.}
    \label{sd001}
\end{figure}

We restrict ourselves to the study of the 3D families bifurcated from x1 as
stable, since they are expected to attract around them regular orbits that will
support the thick spiral arms. The way these families are introduced in the
system is usually presented with a stability diagram, which gives the variation
of the stability of the p.o. by means of the two stability indices, $b_1$ (for
the radial perturbations) and $b_2$ (for the vertical perturbations). In        
Fig.~\ref{sd001} we show the stability diagram for model M1. We focus ourselves
at the intersections of the x1 vertical stability index $b_{2}$ with the $-2$
axis. At these points new 3D families, of the same multiplicity with x1, are
introduced in the system \citep{1985CeMec..37..387C}. Usually the $b_2$ index
dives below the b$=-2$ axis for a certain $\Delta E_j$ interval and then it
returns back to values larger than $-2$, giving rise to successive $S\rightarrow
U\rightarrow S$ transitions. The x1v1, x1v3, x1v5 families are bifurcated from
x1 at the $S\rightarrow U$ stability transitions, as stable. 

In order to have a better view of the variation of the stability indices,
Fig.~\ref{sd001} is split in two parts. In Fig.~\ref{sd001}a we give the
stability curves for $-1600\lesssim E_{j} \lesssim -1186.45$ and in
Fig.~\ref{sd001}b for $-1186.45\lesssim E_{j} \lesssim -1022.71$. The indices of
x1 are drawn with black, while its 3D bifurcations are drawn with the colours
used for the same families in Fig.~\ref{cc001}. They are also indicated with
arrows. In the $E_{j}$ intervals in which a family becomes complex unstable we
draw a straight line segment at $b_{1,2}=0$. The vertical red line in figures
\ref{cc001} and \ref{sd001} denotes the location of the 4:1 resonance in model
M1.

\begin{figure}
\epsscale{1.35}
\plotone{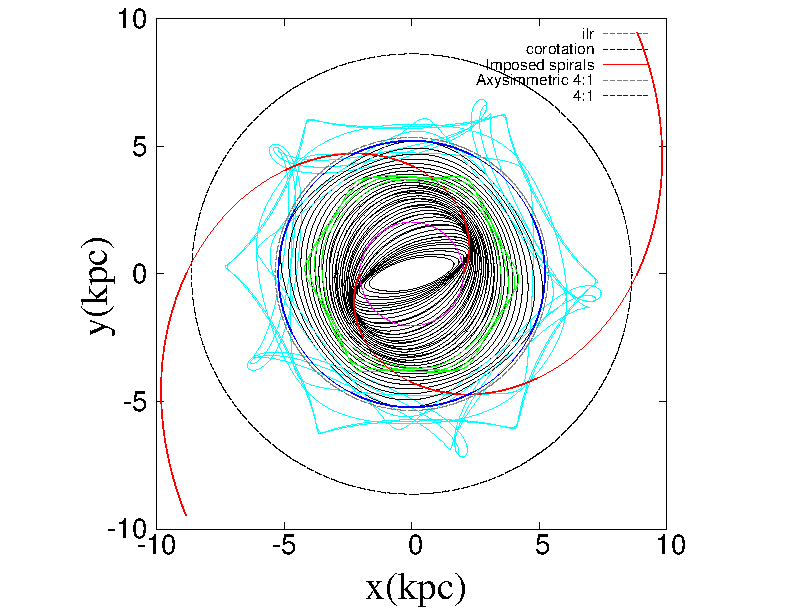}
\caption{Model M1: The superposition of p.o. that could act as building blocks
for the spirals of this model. The p.o. of the x1-tree with elliptical
projections on the equatorial plane are plotted with black. They belong to the
x1, x1v3 and x1v5 families. 2D p.o. orbits beyond the 4:1 resonance are plotted
in light blue, while with green color we plot x1v3 orbits with polygonal
projections on the equatorial plane. The magenta dashed line defines the ILR of
the model, while the gray and blue dashed lines the location of the 4:1
resonance for the axisymmetric and full potential, which for M1 practically
coincide. Finally the black dashed line indicates corotation. The imposed
potential minima of the spiral are given with a red solid line. The system 
rotates clock-wise.}
\label{sp001}
\end{figure}

We used these p.o. in order to find a reasonable orbital backbone that could 
support our imposed 25$^{\circ}$ bisymmetric spiral. However, as we can observe 
in Fig.~\ref{sp001}, such a backbone is not provided by model M1. In 
Fig.~\ref{sp001}  the p.o. belonging to the x1-tree (x1, x1v3, x1v5) are plotted 
in black. Firstly we observe that inside the radial 2:1, Inner Lindblad 
Resonance (ILR), denoted with a magenta circle, the members of the x1-tree p.o., 
belonging to x1 and x1v1 families, are elongated red ellipses. 
Despite the fact that the arms 
of the spiral potential end at ILR, at $r=2.03$~kpc, the particles inside this 
radius feel an m=2 component, due to the presence of the arms. 
These ellipses have their apocentra almost along an axis.
Thus, they could support, a non-imposed, central bar-like component in the central parts of 
the model. 

Beyond ILR, we find orbits with elliptical projections on the equatorial plane
up to the radial 4:1 resonance. However, their apocentra are not aligned with
the imposed spiral. They tend to form a tighter and weaker one. The 2D families
beyond 4:1, depicted in light blue in Fig.~\ref{cc001}, have polygonal shapes
with apocentra not aligned with the imposed spirals. They are plotted again in
light blue in Fig.~\ref{sp001}. We find also 3D p.o, which have a polygonal
shape in their projections on the equatorial plane. These projections appear at
smaller radii than the 2D orbits for the same energy, since they have also a
certain thickness. Such orbits, belonging to the x1v3 family are plotted with
green colour in Fig.~\ref{sp001}. It is evident that the overall response of the
model does not provide a skeleton of stable p.o. with which we could proceed in
building a spiral with regular orbits at any region and between any resonances
of the model.
  
\subsection{Model $M_{s}/M_{d}=0.04$}\label{m4}
\begin{figure}
\epsscale{1.15}
\plotone{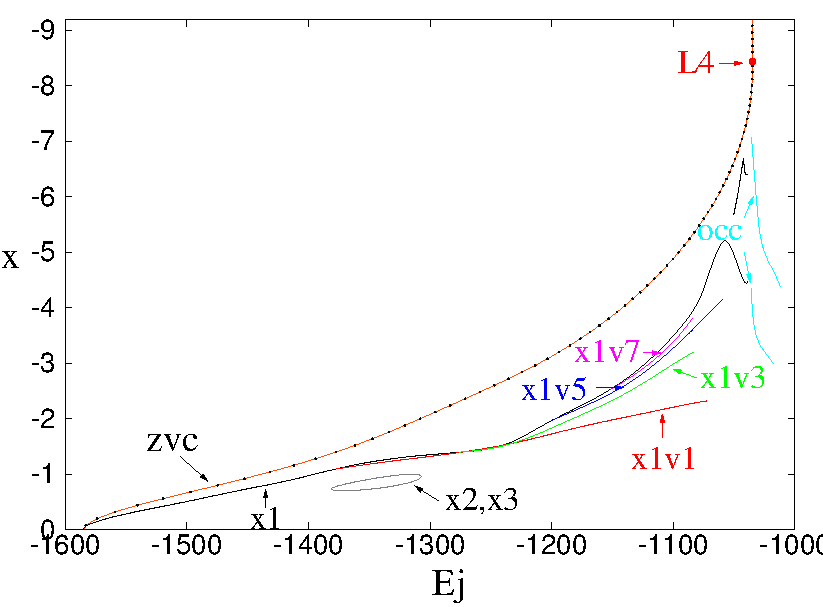}
\caption{Model M4: Characteristic curves in the $(E_j,x_0)$ plane. We use
the same colors as in Fig.~\ref{cc001} for plotting the corresponding families
of the x1-tree. In addition we give in magenta the x1v7 3D family, while the
loop in the lower left part of the figure is the characteristic curve of the
x2, x3 pair.}
    \label{cc004}
\end{figure}

\begin{figure}
\epsscale{1.15}
\plotone{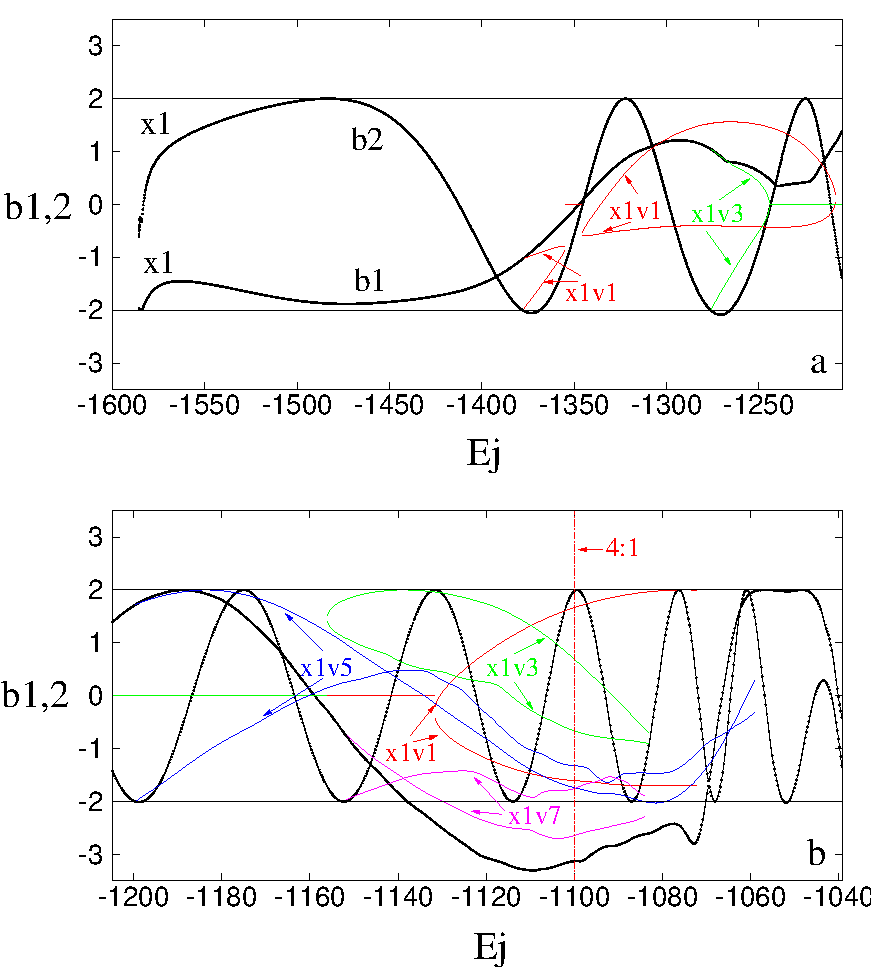}
\caption{Model M4: (a) Stability curves for x1 and for the main 3D families of
the x1-tree in the range $-1600\le E_j\le -1204.8$. The radial index is $b1$,
while the vertical $b2$. (b) The same indices in the range $-1204.8\le E_j\le
-1039.3$. Colors as in Fig.~\ref{cc004}. The dashed vertical red line, labeled
``4:1'', indicates the $E_j$ at which the x1 p.o. start having a rectangular
morphology.}
    \label{sd004}
\end{figure}

The next model we present has $M_{s}/M_{d}=0.04$ and we will refer to it as
model M4. In Fig.~\ref{cc004} we give for M4 the characteristic curves of the
families of the x1-tree, projected in the $(E_j,x_0)$ plane. Colors are as in
Fig.~\ref{cc001} and arrows point to each family. For this model we give also
the loop of the x2, x3 characteristic, while one more 3D family, x1v7, is
included, drawn in magenta color. In Fig.~\ref{sd004} we present the stability
diagram for the x1-tree families of this model. We keep the same colors for the
families as in Fig.~\ref{cc004}. In Fig.~\ref{sd004}b, the $E_j$ for which the
x1
orbits start having a rectangular character is indicated with a dashed vertical
line as ``4:1''. We indicate this $E_j$ also as ``4:1'', since it is the shape
of the orbits that acts as obstacle in the continuation of the support of a
spiral pattern along the loci of the spiral potential by the periodic orbits of
the system. This way we have an ``effective'' 4:1 resonance at a smaller
energy than the one predicted for the 4:1 location from the axisymmetric part
of the potential.

% \begin{figure} \epsscale{1.15} \plotone{cc004_def3d.png} \caption{Model M4: The
% same as in Fig.~\ref{cc004} but in the $(E_j,x_0,p_{x_0})$ projection. The 3D
% families have additional $z_0$ and $p_{z_0}$ non-zero initial conditions.}
% \label{cc0043d} \end{figure}

\begin{figure}
\epsscale{1.35}
\plotone{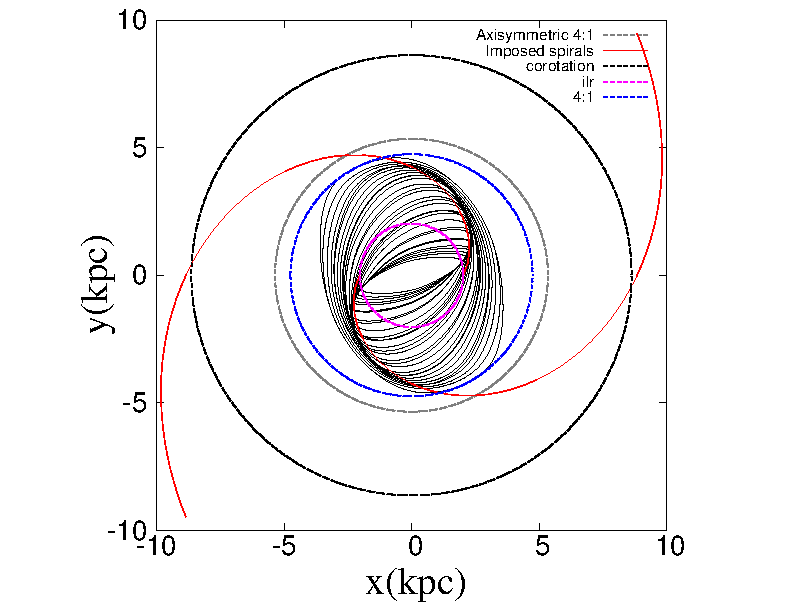}
\caption{Model M4: Superposition of x1 and x1v3, x1v5, x1v7 
projections of p.o. on the equatorial plane that build a spiral-enhancing 
backbone. All orbits are narrower than 0.3~kpc. We observe that they do not 
exceed the region of the 4:1 resonance (circles and colors of orbits, as in 
Fig.~\ref{sp001}. The blue circle at the 4:1 resonance now indicates an
``effective'' 4:1 resonance - see text).}
    \label{sp0041}
\end{figure}

\begin{figure}
\epsscale{1.35}
\plotone{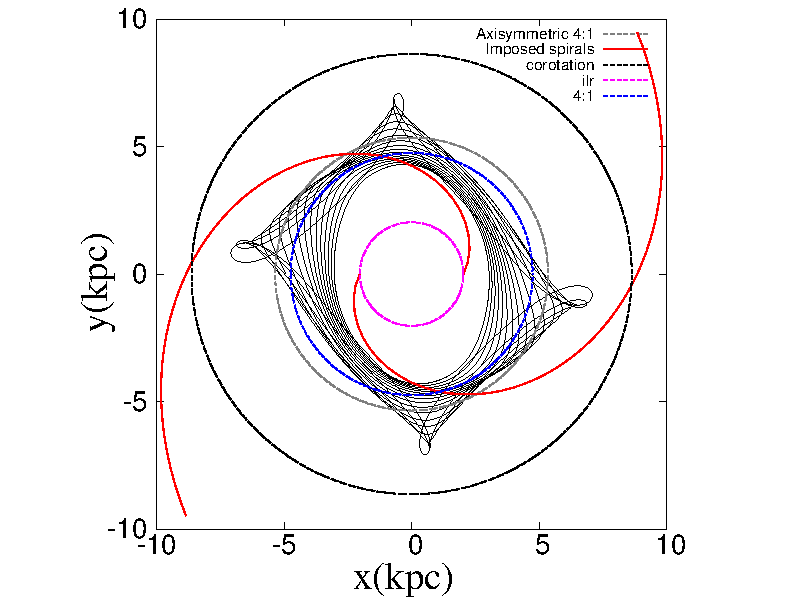}
\caption{Model M4: A set of x1 p.o. close, beyond the 4:1 
resonance. Their 
shape and orientation support a bifurcation of the main arms
(circles as in Fig.~\ref{sp0041}).}
    \label{sp004not}
\end{figure}

In Fig.~\ref{sp0041} we isolated x1-tree p.o. that offer a backbone for 
supporting the $25^{\circ}$ spiral of our model. It includes x1, as well as 
projections on the equatorial plane of p.o. belonging to the families x1v3, x1v5 
and x1v7. In the case of x1, the last spiral supporting orbit was at energy 
$E_j\approx-1100$. Beyond that energy the morphology of the x1 orbits was 
rectangular-like. This is the $E_j$ value indicated with an arrow in 
Fig.~\ref{sd004}. The circles drawn in Fig.~\ref{sp0041} show the locations of 
the ILR (innermost), the location of the 4:1 region as defined by the appearance 
of rectangular-like x1 p.o., the 4:1 radius as expected from the axisymmetric 
part of the potential and corotation (outermost). We observe that the included 
p.o. of the x1-tree offer a backbone that tends to support the spiral (plotted 
with red solid line) up to the 4:1 resonance. The x1 p.o. 
close to 4:1, but at a slightly higher $E_j$ do
not help the spiral extending further out (Fig.~\ref{sp004not}). Contrarily, 
they tend to support density maxima off the spiral (red curve) that can be 
described as bifurcations of spiral arms with a pitch angle smaller than 
$25^{\circ}$, as well as a boxy feature in the middle of the way to corotation. 
All these are conspicuous in Fig.~\ref{sp004not} that shows the orbital backbone 
offered by x1 p.o. for $E_j$'s larger than $E_j\approx-1100$. 

The projections of the 3D families in Fig.~\ref{sp0041} are also subject to a
morphological evolution towards rhomboidal shapes as their $E_j$'s increase.
However, simultaneously, they gain in height, causing their rhomboidal
projections to appear at smaller radii in general, than the x1 p.o. Taking into
account the morphological evolution of all orbits involved in the building of an
orbital skeleton of p.o. that supports a thick spiral, we find that their
thickness at the 4:1 resonance region ($r\approx$ 5~kpc) is about 0.3~kpc. The
corresponding isodensity contour of the axisymmetric part of the model will be
given in subsequent figures with orbital edge-on profiles. This is in agreement
with a reasonable geometry for an Sc galaxy. The last 3D spiral enhancing p.o.
are found at $E_j\approx -1179.523$ for the x1v3 family, at  $E_j\approx
-1143.313$ for x1v5, and at $E_j\approx -1100$ for x1v7 (close to the last x1
spiral-enhancing p.o.). We will return to this point in Section~\ref{sec:respo}.

\begin{figure}
\epsscale{1.35}
\plotone{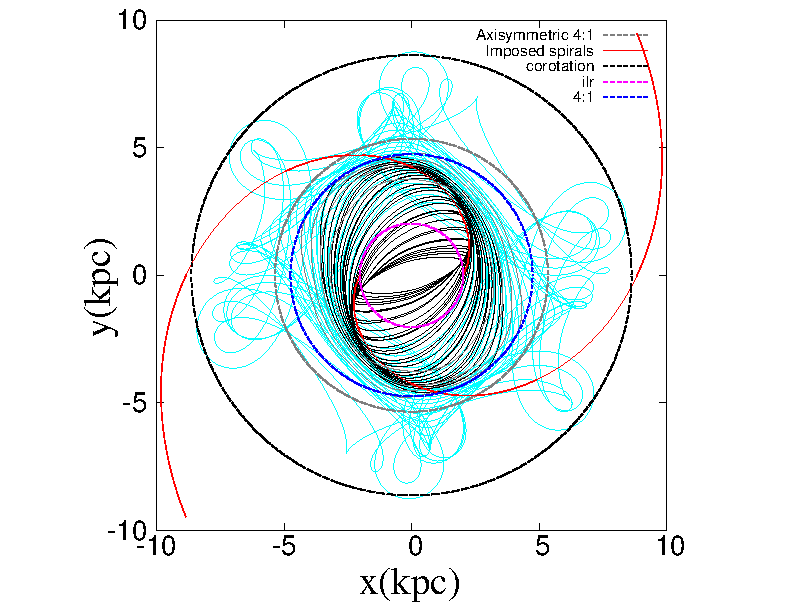}
\caption{Model M4: A sample of p.o. belonging to all 2D and 3D families found 
in the system. The 2D families are in the equatorial plane and the 3D are 
projected on it. The orbits that have shapes and orientation supporting the 
spiral pattern (red line) are plotted with black, while the rest in light-blue 
color. The black orbits reach the 4:1 resonance. The circles indicate 
resonances as in Fig.~\ref{sp0041}.}
    \label{sp0042}
\end{figure}

If we consider all p.o. discussed so far and plot them in a 
single figure 
we
end up with Fig.~\ref{sp0042}. The p.o. that provide a skeleton for supporting
the spiral pattern are plotted in black, while the non-supporting in light blue.
It is clear that on the equatorial plane of the model there are 
p.o. to
support the spiral pattern up to the 4:1 resonance region. It is not obvious at
all that the light-blue \textit{periodic}-orbits could provide a backbone that
would help the pattern continue towards corotation.

\begin{figure}
\epsscale{1.2}
\plotone{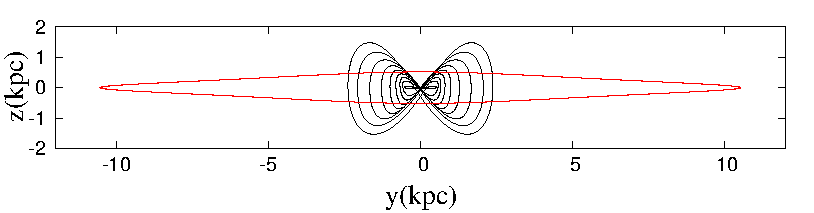}
\caption{The $(y,z)$ projection of a set of x1v1 p.o. for model M4. These are 
orbits that extend inside the ILR of the model, do not support the spiral and 
form a peanut-shaped, bar-like component. The red curve is an 
isodensity contour that refers to
the 
axisymmetric potential.} The spiral supporting orbits of the previous figures 
do 
not exceed the height of this isocontour.
    \label{eoyxv1004}
\end{figure}

The inner limit of the spiral structure is by construction at the ILR. Appart
from the x1 orbits, in all previous figures the 3D families we discussed are the
x1v3, x1v5 and x1v7, i.e. stable orbits bifurcated from x1 at the vertical 3:1,
4:1 and 5:1 respectively. The family x1v1, which seems to be very important for
the edge-on profiles of barred galaxies \citep{2002MNRAS.337..578P} has
elongated elliptical projections, similar to the x1 p.o. inside the ILR. In the
PERLAS models its orbits do not contribute to the reinforcement of the spiral.
They support a bar-like structure in the central parts despite the fact that
PERLAS does not include an explicit central bar component. This bar-like
response of our non-barred, normal spiral model was found in all models we
examined. Thus the PERLAS spiral keeps its 25$^{\circ}$ pitch angle only beyond
the inner 2:1 resonance. Crossing this resonance towards the center of the
galaxy the pitch angle opens and the response has been always bar-like.
Interesting is also to observe the $(y,z)$ edge-on view of a set of x1v1 orbits.
We realize that the shape of the orbits supports a peanut-shaped profile
(Fig.~\ref{eoyxv1004}). The y-axis 
is not aligned with the major axis of the bar. The $\infty$-shape 
figure 
observed in Fig.~\ref{eoyxv1004} appears in this particular projection. 
 
%\subsection{Model $M_{s}/M_{d}=0.05$}
Models with $0.03 \lessapprox M_{s}/M_{d} \lessapprox 0.05$ give similar 
results with those discribed for model M4. We only observe that the x1 
p.o. become rectangular or rhomboidal-like at an earlier energy, $E_j$, which 
deviates more from the location of the axisymmetric 4:1 resonance. 
 
% In Fig.~\ref{sp005_02} we present again representatives of all calculated 
% families of p.o. this time for model M5, which has $M_{s}/M_{d}=0.05$. It is the 
% corresponding to Fig.~\ref{sp0042} for model M4. The two models provide similar 
% orbital backbones with periodic orbits, with the only difference that the 
% rhomboidal in shape x1 p.o. appear at an earlier $E_j$ with respect to the 
% energy of the axisymmetric 4:1 resonance, than M4. 

% \begin{figure}
% \epsscale{1.35}
% \plotone{sp005_02def2.png}
% \caption{The same as Fig.~\ref{sp004}, this time for model M5 with 
% $M_{s}/M_{d}=0.05$. The circles indicate 
% resonances as in Fig.~\ref{sp001}. }
%     \label{sp005_02}
% \end{figure}

\subsection{Model $M_{s}/M_{d}=0.07$}
\begin{figure}
\epsscale{1.15}
\plotone{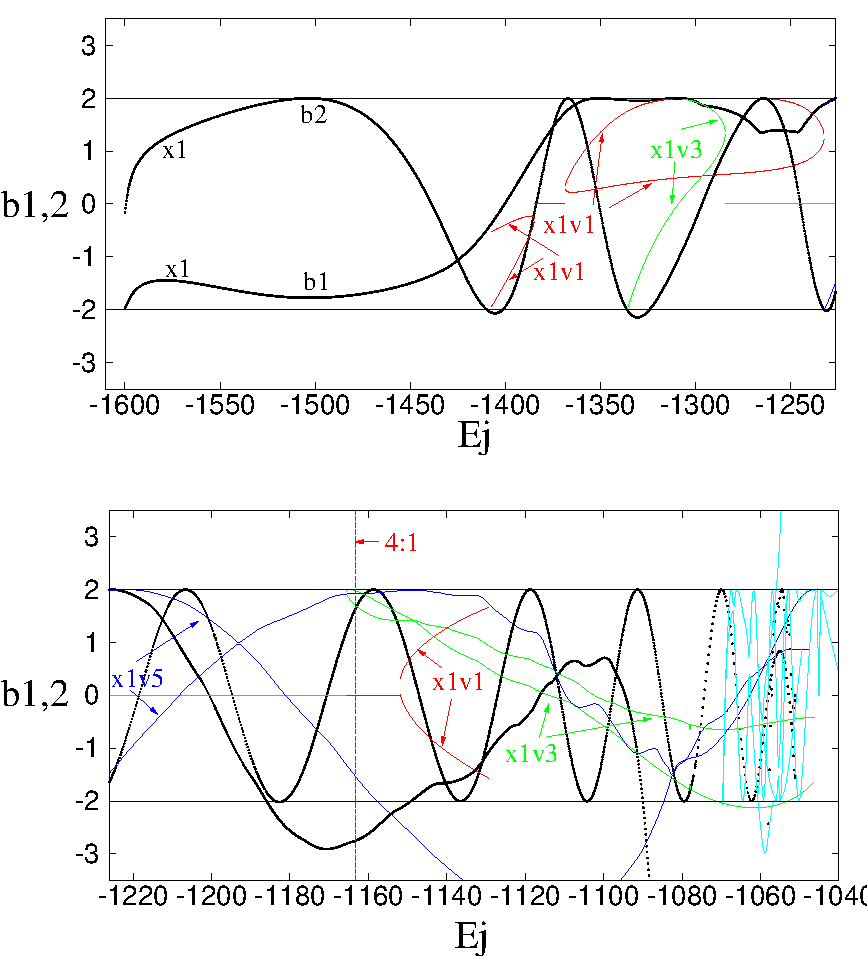}
\caption{Model M7 with $M_{s}/M_{d}=0.07$. (a) The evolution of the stability
indices in the range $-1600 \lessapprox E_j \lessapprox -1230$. (b) The
stability indices in the range $-1230 \lessapprox  E_j \lessapprox -1040$. The
families are plotted with the same colors as in Fig.~\ref{cc004}. The dashed
vertical red line indicates the $E_j$ at which the x1 p.o. become rhomboidal.}
    \label{sd007}
\end{figure}

The orbital behavior of our models started changing when we reach the ratio 
$M_{s}/M_{d}=0.07$, which is our model M7. As we can observe in 
Fig.~\ref{sd007} that shows the evolution of the stability indices, the 
vertical resonances appear now shifted to smaller energies than in the previous 
models. Families x1v1, x1v3 and x1v5, introduced as stable in the 
vertical 2:1, 3:1 and 4:1 resonances respectively, have larger complex unstable 
parts, compared to models with smaller $M_{s}/M_{d}$ ratios.

\begin{figure}
\epsscale{1.35}
\plotone{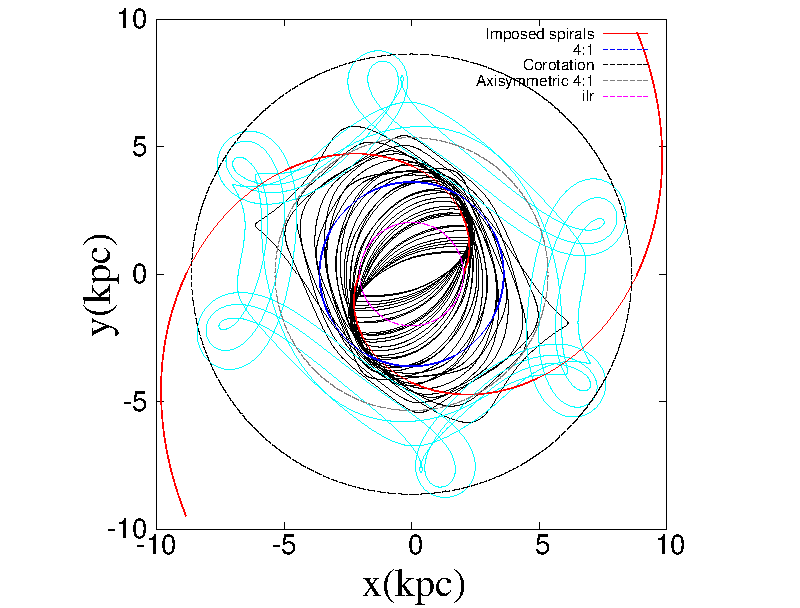}
\caption{Model M7. Periodic orbits of the basic 2D and 3D families of the model 
projected on the equatorial plane. Plotted circles and colors of orbits are as 
in Fig.~\ref{sp001}. The apocentra of the x1 p.o. up to the 
4:1 resonance} are now slightly ahead of the 
spiral of the model in the sense of rotation.
    \label{sp007_01}
\end{figure}

The backbone of p.o. for model M7 is given in Fig.~\ref{sp007_01}. We plot again
2D orbits of the planar families and representatives of the stable 3D families.
Apart from the fact that the x1 orbits become rhomboidal at a shorter distance
from the center than before, we observe that the candidate periodic orbits to
support the spirals have their apocentra ahead, in the sense of rotation, of the
potential minima (indicated with a continuous red line). Nevertheless, they are
not far away from them.

\vspace{0.5cm}
\section{Contribution of quasi- and non-periodic orbits}\label{sec:orbitals}
The comparison of the backbones of p.o. we calculated for models with $0.01 \leq
M_{s}/M_{d} \leq 0.1$ led to the conclusion that the best choice for building a
$25^{\circ}$ spiral is in the range $0.03 \leq M_{s}/M_{d} \leq 0.07$. As a
typical case we have chosen model M4 to present the contribution of quasi- and
non-periodic orbits to the modeled spiral pattern. 

\subsection{Orbits in model M4}
\begin{figure}
\epsscale{1.2}
\plotone{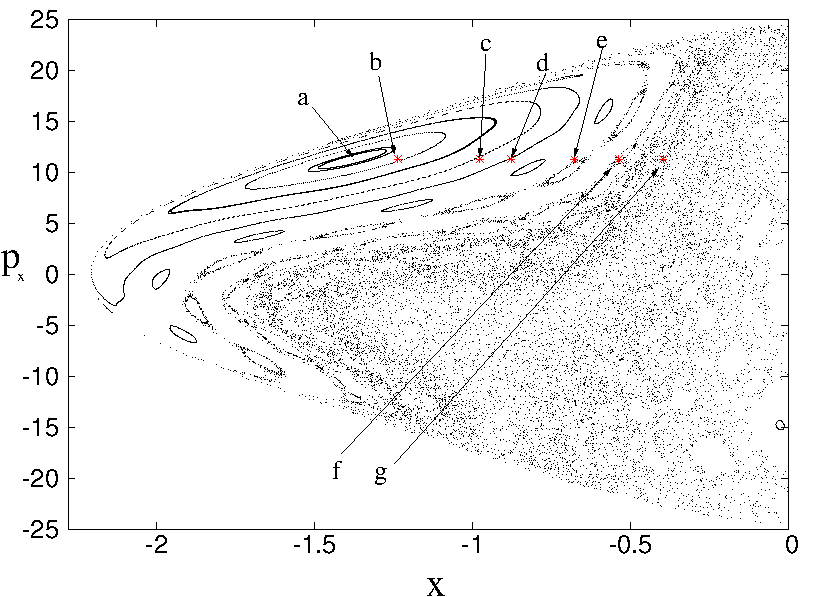}
\caption{The Poincar\'e surface of section at $E_j=-1286.277$ for model 
$M_{s}/M_{d}=0.04$. The arrows indicate the position
    of the orbits shown in figure \ref{or12862777}}
\label{pss1286277}
\end{figure}

\begin{figure*}
\epsscale{1.0}
%\plotone{or12etc.png}
\plotone{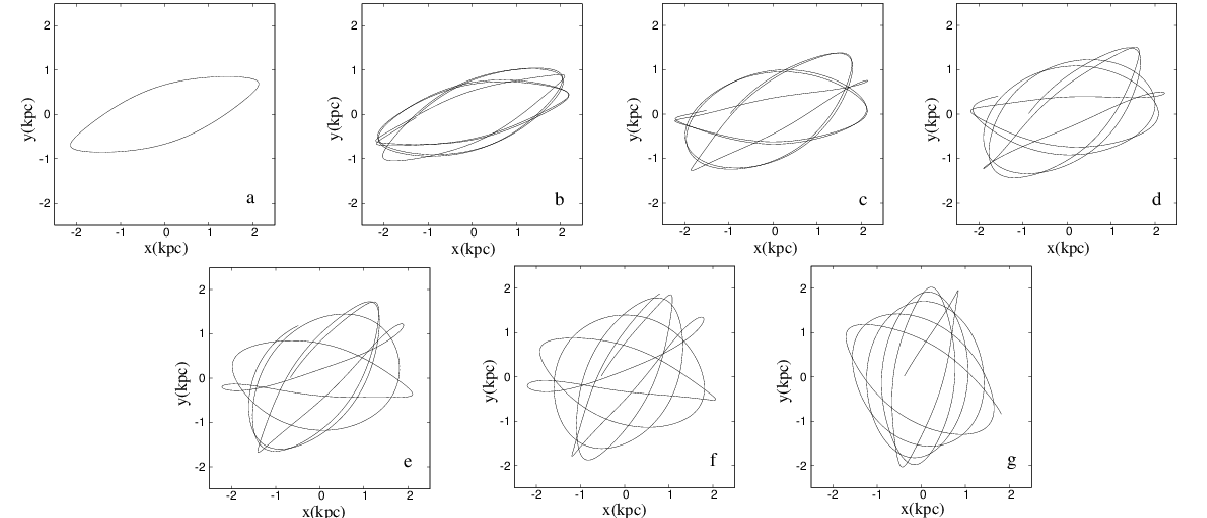}
\caption{Planar orbits of M4 are given for $E_j=-1286.277$. (a) The x1 p.o.  
From (b) to (f) are given orbits with initial conditions deviating from x1 by 
$\Delta x =$ 0.14~kpc (b), 0.4~kpc (c), 0.5~kpc (d), 0.7~kpc (e), 0.84~kpc (f) 
and 0.98~kpc (g). All orbits are integrated for 5 x1 orbital periods. The 
locations of the initial conditions of the orbits on the 
Poincar\'e surface of section
are indicated in Fig.~\ref{pss1286277}.}
\label{or12862777}
\end{figure*}

For this purpose we chose a typical $E_j$ that harbors a stable x1 p.o. from the
backbone of the M4 model spiral. The $(x,p_x)$ Poincar\'e surface of section at
$E_j=-1286.277$ for planar orbits in the case of model M4 is presented in
Fig.~\ref{pss1286277}. The location of the x1 p.o. on the surface of section is
indicated with the arrow labeled with ``a'' and its morphology is given in
Fig.~\ref{or12862777}a. The location of the rest of the orbits depicted in
Fig.~\ref{or12862777} is indicated with arrows in the surface of section in
Fig.~\ref{pss1286277}. The letter that characterizes an orbit on the surface of
section corresponds to the label of each panel in  Fig.~\ref{or12862777}. Their
initial conditions deviate from those of x1 by $\Delta x$ = 0.14, 0.4, 0.5, 0.7,
0.84 and 0.98~kpc for orbits in (b), (c), (d), (e), (f) and (g) respectively.
All orbits have been integrated for five x1 orbital periods. As we can observe
in Fig.~\ref{pss1286277}, orbits (b), (c) and (d) belong to invariant curves
around x1, orbit (e) to an orbit on the chain of islands that follows, while
orbits (f) and (g) are chaotic orbits. Orbit (f) is chosen to be at the edge of
the stability island and orbit (g) well inside the chaotic sea. By comparing the
size, the morphology and the orientation of these orbits with respect to the
loci of the 25$^{\circ}$ spiral, we realize that those in 
Fig.~\ref{or12862777}b to e, support during their integration the enhancement of
the density of the model at the spiral arms region. Orbit (f) supports the
spiral only partly, while orbit (g) does not.

\begin{figure*}
\epsscale{1.2} 
%\plotone{leo_fig15def.png}
\plotone{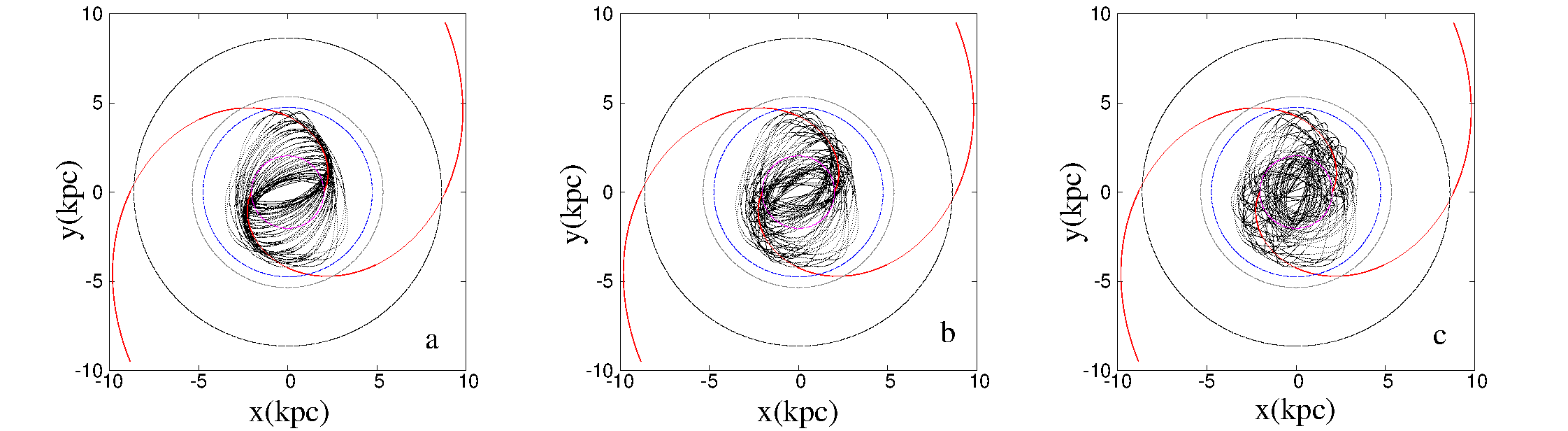}
\caption{M4: The set of x1 p.o. used in Fig.~\ref{sp0041} perturbed by 
$\Delta 
x=$0.14, 0.5 and 0.98~kpc from (a) to (c) respectively. The 
larger perturbed orbits reinforce less the imposed spiral, which is plotted in
red. Circles as in Fig.~\ref{sp0041}.}
\label{sp004per02}
\end{figure*}

As a next test for estimating the contribution of the 2D quasi- and non-periodic
orbits to the enhancement of the spiral structure we applied the same $\Delta x$
perturbations successively to the set of x1 p.o. used in Fig.~\ref{sp0041}. 
These p.o, each at a different $E_j$, are typical orbital building
blocks of
the spirals. They have the right orientation and shape, so
that they enhance the spirals, due to the relative position of their apocentra
with respect to the potential minima of the model. The result is presented in
Fig.~\ref{sp004per02}. The x1 p.o. in Fig.~\ref{sp0041} have been perturbed by
$\Delta x =$~0.14, 0.4, 0.5, 0.7, 0.84 and 0.98~kpc and three of the
resulting sets of
non-periodic orbits are given in Fig.~\ref{sp004per02}a to
\ref{sp004per02}c. Namely, we plot the orbits perturbed by $\Delta x
=$~0.14,
0.5 and 0.98~kpc
respectively. We realize that 
the orbits contribute less as we deviate from the initial conditions of the p.o.
In all orbits of the specific sample we use to describe this property, as well
as in many other cases of orbits we have examined, the quasi-periodic orbits 
around x1
had the largest contribution to the
imposed spiral.  The enhancement of the spirals from orbits at the borders of
the stability islands on the surfaces of section, or from chaotic orbits, like
the one in Fig.~\ref{sp004per02}c, was less evident.

\begin{figure*}
\epsscale{1.2}
%lotone{sp004perdz.png}
%\plotone{leo_fig16def.png}
\plotone{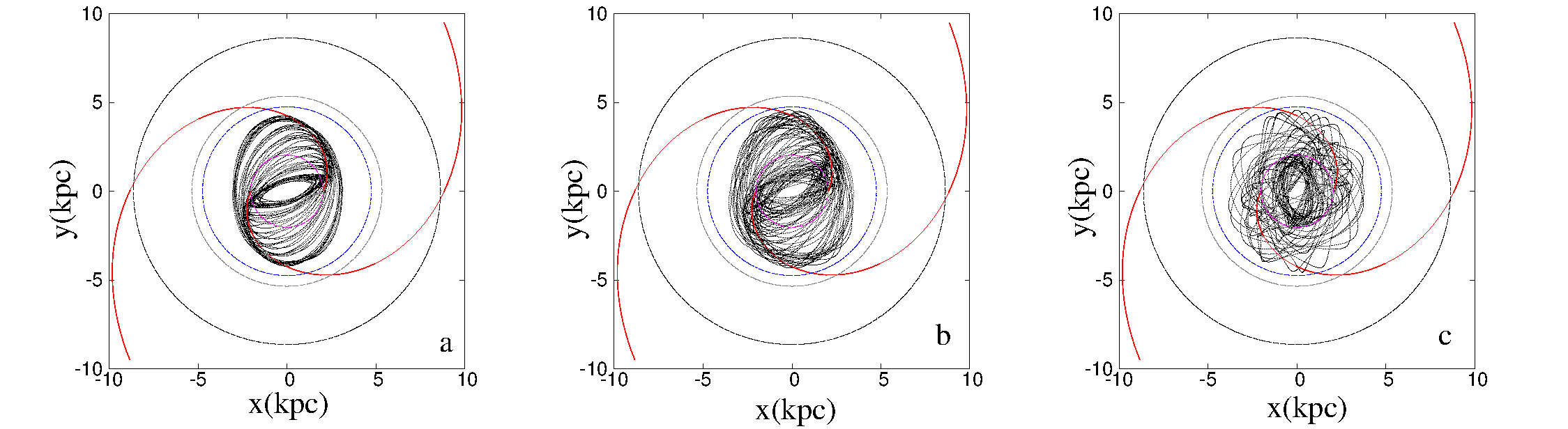}
\caption{M4: The spiral pattern of the orbits in Fig.~\ref{sp004per02} 
perturbed
in the $z$ direction. The orbits in (a) and (b) support the spiral pattern when
projected on the equatorial plane of the model. These orbits do not exceed a
height of 0.3~kpc, away from the $z=0$ plane. The orbit in panel (c) does not 
reinforce the spiral
pattern.}
     \label{sp004perdz}
\end{figure*}

Additional perturbations have been applied to radially perturbed 
p.o. like
those discussed in the previous paragraph, this time in the vertical
direction. Typical cases are
presented in Fig.~\ref{sp004perdz}, which are the orbits depicted in
Fig.~\ref{sp004per02} perturbed also in the $z$ direction. Initially the 
radially
perturbed p.o. orbits have $z_0=0$. We
started increasing the $z$ coordinate of the orbit and we have checked their
projection on the equatorial plane. We stopped increasing $z_0$ as long as the
projection did not support the spiral any more. The edge-on profiles of these
sets of spiral-supporting orbits were similar to those we 
obtained when we have considered
just the corresponding
periodic orbits. Namely, the thickness of the profiles did not 
exceed $z=$0.3~kpc above or below
the equatorial plane at
$r\approx$5~kpc, while we have $z\approx 0.54$ kpc at $r=0$. The orbits
in
panels Fig.~\ref{sp004perdz}a and Fig.~\ref{sp004perdz}b support the spiral.
However, all perturbations in the $z$
direction we tried for the orbit in Fig.~\ref{sp004per02}c gave orbits that do
not reinforce the
spiral pattern. We note that by perturbing a chaotic orbit, this does not
necessarily imply that the perturbed orbit will be chaotic as well, since by the
displacement of the initial conditions we may reach regions of the phase space
occupied by tori.  

%%%%%%%%%%%%%%%%%%%%%%%%%%%%%%%%%
%%LATER
%\begin{figure}
%\epsscale{1.2}
%\plotone{pss1043_385.png}
%\caption{The Poincar\'e surface of section for model $M_{s}/M_{d}=0.04$ at
%$E_j=-1043.385$. The left panel
%    shows the invariant curves of the family that supports the spirals between 4:1 and corotation.}
%    \label{pss00401}
%\end{figure}
%%%%%%%%%%%%%%%%%%%%%%%%%%%%%%%%%%

% \begin{figure*}
% \epsscale{1.2}
% \plotone{eoy004per.png}
% \caption{M4: The edge on profiles of the models shown in 
% Fig.~\ref{sp004perdz}. We observe that they have narrow, cylindrical shapes.}
% \label{eoy004per}
% \end{figure*}

\begin{figure}
\begin{center}
 \includegraphics[scale=0.33]{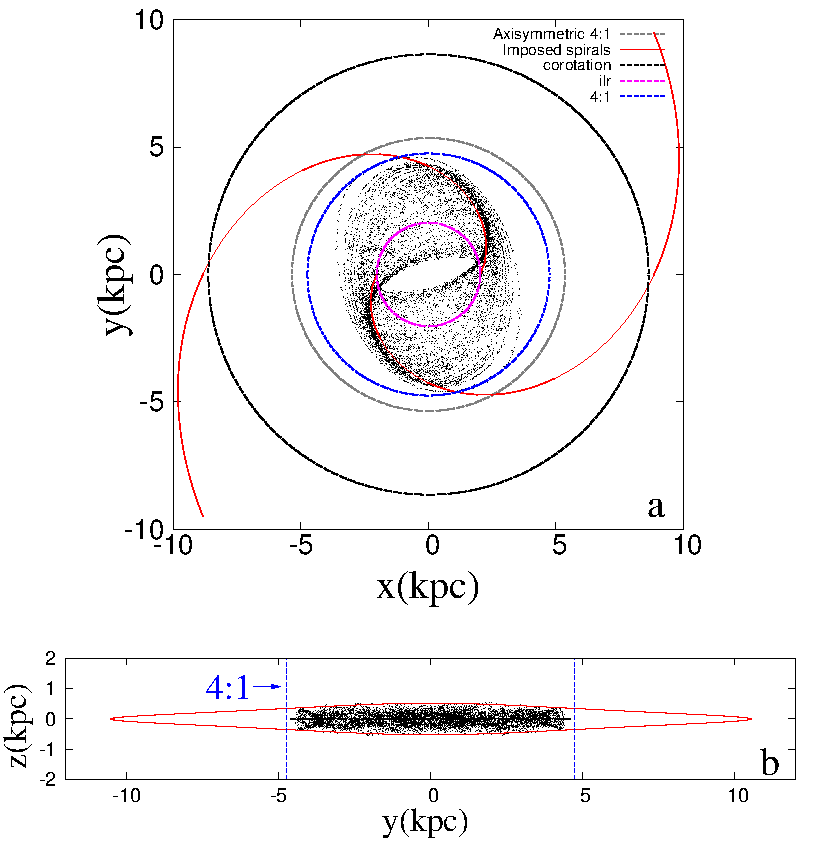}\\
\end{center}
    \caption{M4: (a) The face-on view of a set of regular orbits trapped around 
x1, x1v3, x1v5 and x1v7 p.o. These orbits support a thick spiral between the 
radial 2:1 and 4:1 resonances on the $z=0$ plane (circles as in 
Fig.~\ref{sp0041}). (b) The edge on
profile of the spiral pattern. The plotted in red isodensity contour
refers
to the axisymmetric part of the imposed potential. The
height of the perturbed orbits does not
exceed $0.3$~kpc away from the $z=0$ plane at the radial 4:1 resonance region.}
    \label{sp004per01}
\end{figure}

Along the largest parts of the spiral of the models, besides the stable x1 p.o.,
exist also stable 3D families of periodic orbits. These orbits will trap again
around them regular orbits and it is expected that they will also participate in
the enhancement of the density of the spirals. For this 
purpose we apply perturbations to all
spiral supporting p.o., 2D and 3D, depicted in Fig.~\ref{sp0041}. By doing so,
we can find a thick spiral pattern supported by 3D regular orbits up to the
radial 4:1 resonance, at the region where the x1 p.o. become rhomboidal. The
face-on view of this spiral pattern is presented in Fig.~\ref{sp004per01}a and
its edge-on view in Fig.~\ref{sp004per01}b. In Fig.~\ref{sp004per01}b we
plot
also an isodensity contour of the axisymmetric part of the potential that does
not exceed the 0.3~kpc height above or below the equatorial plane at the 4:1
resonance distance. We observe that the total edge-on profile of the set of
orbits is included within this contour.

By perturbing the p.o. of the backbones of the models we investigated we find 
similar results for the cases $0.03 < M_{s}/M_{d} < 0.07$. The 3D spirals are 
reinforced by orbits similar to those we described in the previous paragraphs 
for model M4. In the cases with $M_{s}/M_{d} < 0.03$ the ellipticity of the x1
p.o. and that of the projections of its vertical bifurcations on the equatorial
plane is small. Considering quasi-periodic orbits around them does not improve
the relation between the orbits of the model and the imposed spiral. For
$M_{s}/M_{d} > 0.07$, apart from the large complex unstable parts of the 3D
families the size of the stability islands around x1 decreases, which
makes the support of the spiral structure problematic.

\vspace{1cm}
\section{Response models}\label{sec:respo}
Having studied the basic orbital dynamics, we investigate further
the overall dynamics of PERLAS models by integrating sets of $4.6\times 10^4$
initial conditions. We start by distributing randomly test particles in a disk
of 13~kpc radius in order to obtain an initially homogeneously populated disk.
The velocities of these particles correspond to those of circular motion in the
axisymmetric part of the potential ($M_{s}=0$). Then we increase linearly
$M_{s}/M_{d}$ within two system rotations reaching at the end of this period the
$M_{s}/M_{d}$ ratio that characterizes each model. Following this procedure, the
particles obtain the initial conditions with which they will be integrated in
the final full potential. We follow the response of the model to the imposed
potential for about 30 system rotations. Besides the 2D responses, we calculated
3D response models as well, by placing randomly the particles in cylindrical
configurations, which had a 13~kpc radius and a height of 0.3~kpc above
and
below the equatorial plane. We have
chosen the radius to be the same with the 2D disk, while for the height we have
taken into account the total thickness indicated by the orbital models,
namely
0.6~kpc. This is a height within which the regular orbits can easily support 3D
spirals. The number density of the particles in the initial cylinder is
homogeneous. The initial velocities have been chosen to be those for circular
motion on the equatorial plane at the same radius. This rather arbitrary choice
is used just as a starting point, which leads to the initial conditions in
the full potential after the transient two pattern rotations period. 

\begin{figure}
\begin{center}
 \includegraphics[scale=0.5]{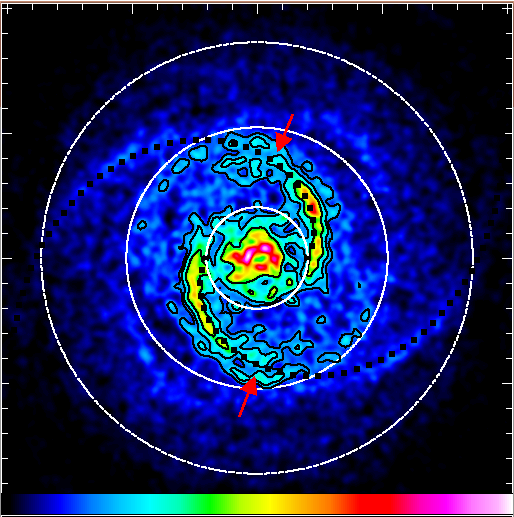}\\
\end{center}
    \caption{The 3D response M4 model after about 25 pattern rotations. The
image is weighted according to the exponential profile of the disk. The circles
indicate the ILR (innermost), 4:1 (middle) and corotation (outermost)
resonances. The red arrows indicate the end of the strong, bisymmetric spiral
pattern. Circles as in Fig.~\ref{sp001}.}
    \label{m4_3dexpo1}
\end{figure}

The snapshots of the response models have been converted to images by means of
the ESO-MIDAS package, taking into account the local number density of the
particles. The results of this exercise pointed again to best matching between
the imposed potential and the response model for the cases 
with $0.03 
\leq M_{s}/M_{d} \leq 0.07$. If we weight the images taking into consideration
that the disk of the imposed potential is exponential with a scale length
3.7~kpc and that the central density of the Schmidt spheroids falls also
exponentially with radius with the same scale length as the disk, we end up to
figures like Fig.~\ref{m4_3dexpo1}, which is our ``density map'' for model M4.
Such images describe the morphology and the relative
importance of the morphological features supported by the orbits of the model.

Fig.~\ref{m4_3dexpo1}, describes the 3D response of model M4, projected on the
equatorial plane. The three drawn circles indicate the ILR (innermost), the 4:1
resonance (middle) and corotation (outermost). The loci of the imposed spirals
are drawn with heavy black dots. The overall response is in agreement with the
results of past studies on the subject based on orbital theory
\citep{1986A&A...155...11C, 1988A&A...197...83C, 1991A&A...243..373P,
1996A&A...315..371P}. Namely, we have a strong, bisymmetric spiral
pattern that starts existing at the ILR region and it is almost in-phase with
respect to the imposed spiral. It broadens and bifurcates just before the radial
4:1 resonance at the points indicated with red arrows. The response morphology
of this model is established already after about 10 pattern rotations. The
pattern depicted in the model, with the two, main, symmetric spiral arms that
are split at larger distances, is frequently encountered in the morphology of
non- or weakly-barred spiral galaxies \citep{2004A&A...423..849G}. 

\begin{figure*}
\begin{center}
 \includegraphics[scale=0.5]{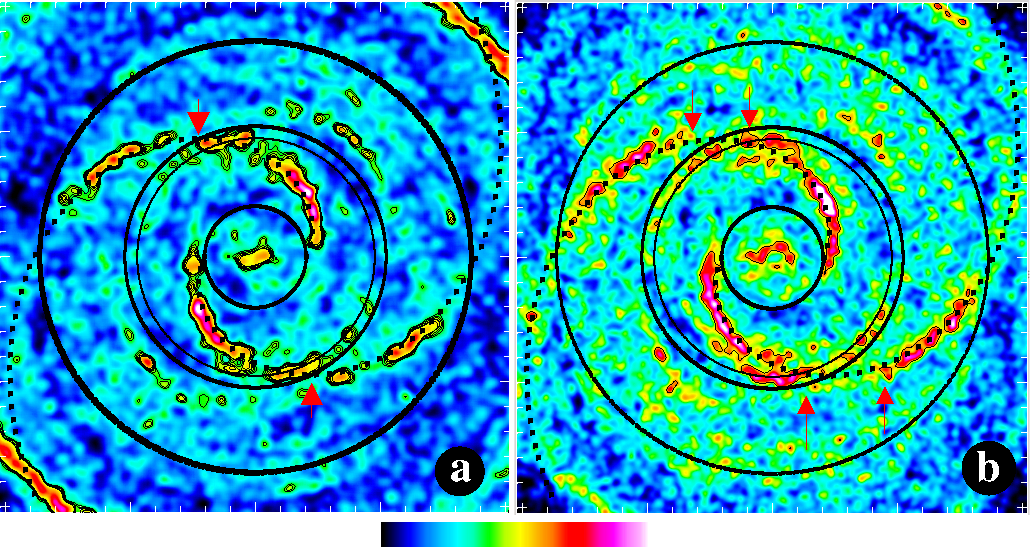}\\
\end{center}
    \caption{Non-weighted for the exponential profile of the disk M4 response
models, after about 25 pattern rotations (a) The response of an infinitesimally
2D disk (b) The response of the corresponding 3D initial configuration (see
text). Circles as in Fig.~\ref{sp0041}.}
    \label{homom4}
\end{figure*}

Faint extensions of the spirals are observed in the region between the 4:1
resonance and corotation. They are faint, since they are located in low density
regions of the disk, due to its exponential profile. In general, extensions of
the spiral arms beyond 4:1 are found in the responses of gaseous models. Typical
examples are given in \citet[][see e.g. figures 3 and 4]{1997A&A...323..762P}.
The difference between those extensions and what we find here, is that now we
have mainly fragments of stellar spiral arms in- or nearly 
in-phase with the imposed
spiral. To better describe their properties we present in Fig.~\ref{homom4} the
images of the 2D and 3D responses of model M4 without weighting them with the
exponential decrease of the disk and spiral densities. Thus, morphological
features appear equally important, regardless of their distance from the center
of the system. In Fig.~\ref{homom4}a it is depicted the response of the 2D,
while in Fig.~\ref{homom4}b that of the 3D
response model (the projection on the equatorial plane). In both panels we
observe a gap at the 4:1 resonance region. However, in both cases we find again
local density enhancements between 4:1 and corotation, close to the imposed
spiral loci of the PERLAS potential. In the 3D case (panel b) the gap is larger
(indicated with two arrows along each arm) but the segments that follow are
longer than the chunks of local density enhancements we find in the 2D model
(panel a). Nevertheless, in neither of the cases we find segments of spiral arms
crossing corotation. 

\begin{figure*}
\begin{center}
 \includegraphics[scale=0.75]{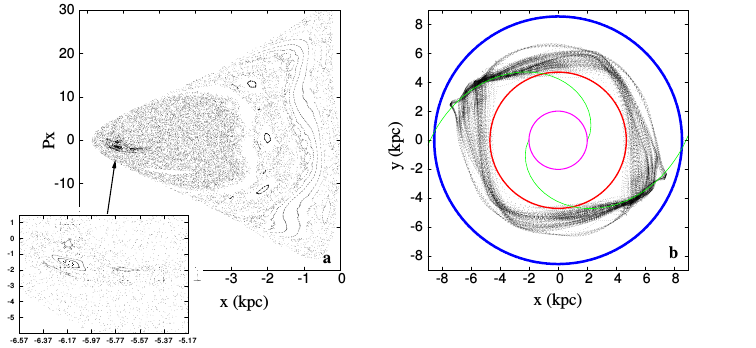}\\
\end{center}
    \caption{The $(x,p_x)$ surface of section of model M4 at $E_j = -1046$. In
(a) we observe tiny stability islands embedded in a chaotic sea. In enlargement
we give the region close to the 4:1 family. Quasi-periodic and sticky orbits
from the region are plotted in (b), together with the spiral loci. We observe
how such orbits can contribute to the enhancement of the spiral arm segments
beyond the gap of the 4:1 resonance (red circle).}
    \label{largej}
\end{figure*}

The orbits of the particles populating the ring between the resonances 4:1 and
corotation have large $E_j$'s. Most of them are in the range $-1056<E_j<-1036$.
A typical $(x,p_x)$ surface of section for $E_j=-1046$ in model M4 is given in
Fig.~\ref{largej}a. We observe that the phase space is dominated by a chaotic
sea. The stable periodic orbits at this $E_j$ do not provide the necessary
backbone for supporting the local density enhancements by regular orbits. Thus
the segments of spirals observed in the models in this region can be only
features reinforced by sticky chaotic orbits. Indeed, the stability islands in
the region are tiny. In Fig.~\ref{largej}a we give in enlargement the area of
the surface of section that includes the stability island of a ``4:1'' family
with rhomboidal orbits (lower right island). As an example, we give in
Fig.~\ref{largej}b quasiperiodic orbits of this family and sticky chaotic orbits
from the region around it, that show how the segments could be supported. The
detailed mechanism, as well as model dependencies, will be investigated in a
subsequent paper. Nevertheless, we note that in this case we have a similar
situation like in the barred-spiral model for NGC~4314, studied by
\citet{1997ApJ...483..731P}, where the sticky chaotic orbits around tiny islands
affect the morphology of the model at the radial 4:1 resonance region.

\begin{figure}
\begin{center}
 \includegraphics[scale=0.45]{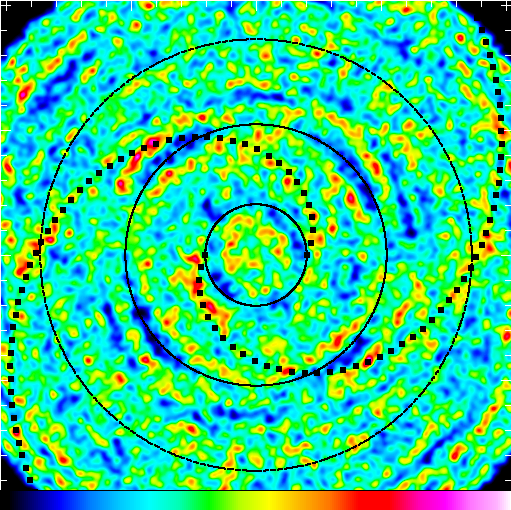}\\
\end{center}
    \caption{Projection on the equatorial plane of the 3D response of model M1
($M_{s}/M_{d}=0.01$) after about 20 patten rotations. No conspicuous spiral arm
structure is discernible.}
    \label{rm01hom}
\end{figure}

At the level of orbital analysis of the present study, it is difficult to name a
specific $M_{s}/M_{d}$ ratio that gives the best matching between imposed
potential and response model. All 2D and 3D PERLAS response models with $0.03
\leq M_{s}/M_{d} \leq 0.07$ have similar overall morphologies. Outside this
range the models show conspicuous deviations from this morphology. For example
the 3D response of model M1 ($M_{s}/M_{d}=0.01$) does not show a well organized
spiral structure. This can be seen in Fig.~\ref{rm01hom} in which we avoid
weighting the image taking into account the exponential profile of the disk.
This way we show that there is an intrinsic discrepancy between the orbits of
the model and the imposed spiral. In the response we do not find  well organized
spiral arms. Only parts of them can found along the imposed spirals. Scattered
arcs of enhanced local densities, in general not along the loci of the imposed
PERLAS spirals, are observed in other regions. This is expected from the
calculated backbone of periodic orbits provided by the model (Fig.~\ref{sp001}).
Nevertheless, the characteristic quadruple gap at the 4:1 resonance is
discernible in Fig.~\ref{rm01hom} along the 4:1 resonance circle. 

\begin{figure*}
\begin{center}
 \includegraphics[scale=0.45]{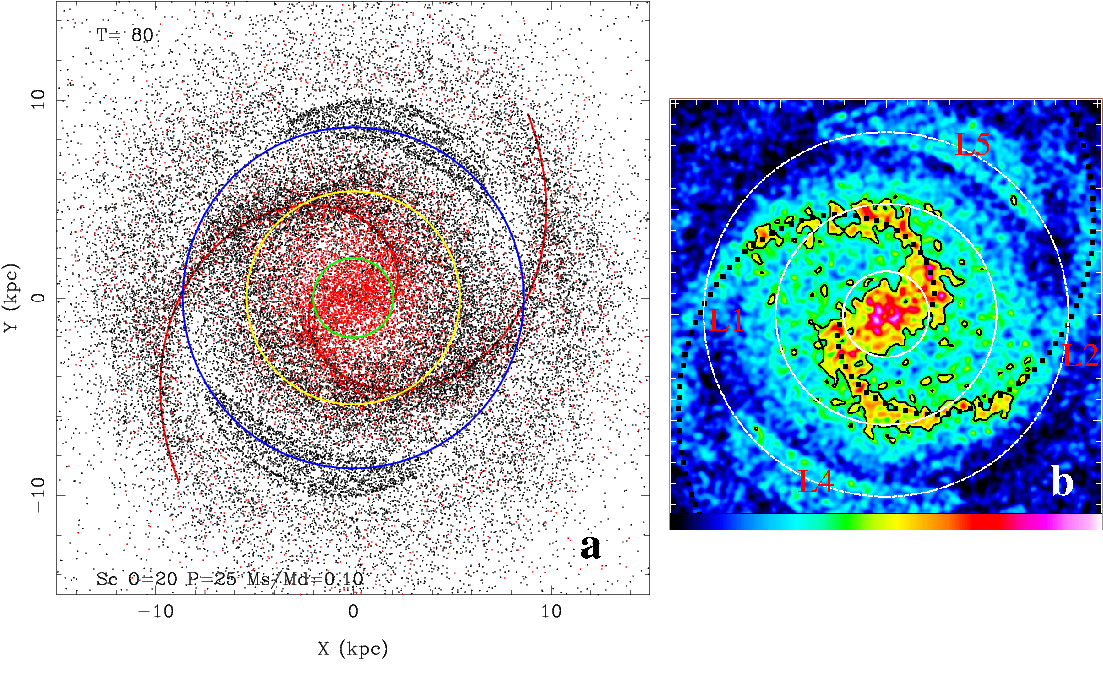}\\
\end{center}
    \caption{Response (3D), of the strong model M10. In (a) we depict
all particles participating in the response model after about 25 pattern
rotations, projected on the equatorial plane. Particles in chaotic orbits are
plotted in red. In (b) we give the image of (a), where we can observe the
irregular shape of the spiral arms, the banana-like regions around L4 and L5
and the beginning of ``chaotic'' spirals starting from the neighborhood f L1
and L2. From the center outwards the circles indicate the ILR, 4:1 and
corotation resonances.}
    \label{m10model}
\end{figure*}

When the mass ratio increases, the alignment of the response density
maxima with the imposed spirals improves and we have the level of agreement we
described for model M4. However, for $M_{s}/M_{d}>0.07$, the sizes of the
stability islands around the x1 p.o. in the $(x,p_x)$ surfaces of sections are
considerably reduced, even at small energies. We find though sticky zones, which
provide to the system sticky-chaotic orbits that behave like regular for several
rotational periods and support a spiral pattern with $25^{\circ}$ pitch angle in
the response models. However, as time increases, the spiral arms become broader
and less well defined. A typical example is given in Fig.~\ref{m10model}, which
is the 3D response model for $M_{s}/M_{d}=0.1$ (hereafter called model M10).
In (a) we present all particles projected on the equatorial plane at time
corresponding to about 25 pattern rotations. We have used the  
GALI$_2$ stability
index \citep{2007PhyD..231...30S, 2008EPJST.165....5S} to characterize the
chaoticity of the orbits and based on this we plot with black the particles on
regular and with red the particles on chaotic orbits. We use the GALI$_2$ value
of the orbit at the time of the snapshot to characterize it and we plot with
black the particles on orbits with  $\log_{10}$(GALI$_2$)$>-8$. However, at
later times the index of an orbit may well fall to smaller values, in which case
the particle will be plotted red. This is reflected in Fig.~\ref{m10model}a,
where we observe that we have more red points closer to the center of the
system, because in this region the system is dynamically more evolved. Gradually
the red region expands towards larger radii. The fact that the region between
the 4:1 resonance and corotation is not red after 25 patter rotations, indicates
that many of the chaotic orbits are sticky, mimicking a regular behavior and
keeping their GALI$_2$ index in the range with $\log_{10}$(GALI$_2$)$>-8$.
Building a spiral pattern by means of such orbits has as a result mainly to
obtain broad spiral arms with fuzzy boundaries in the responses. This can be
observed in Fig.~\ref{m10model}b, where we give the image of model M10,
non-weighted for its exponential character. The response spirals are still
formed close to the loci of the imposed spiral, with a tendency to have their
main part ahead of them, but their boundaries have an irregular character. Also
the characteristic bifurcations of the arms at the 4:1 resonance are much less
conspicuous.

Although models with $0.03 \lessapprox M_s/M_d \lessapprox 0.07$ ratios match
better the imposed spirals than models with $M_s/M_d > 0.07$, the ``strong''
models of our sample show that the role of sticky-chaotic orbits can be in some
cases important for the dynamics of spiral galaxies.

In Fig.~\ref{m10model} we observe two more features that we would like to
discuss. With a spiral forcing $F_{\theta}/F_r \approx 0.175$ at the corotation
region, M10 is the strongest model we have studied. Nevertheless, this forcing
is much smaller than what we encounter in the corotation region of fast, strong
bars \citep{block01}. The banana-like periodic orbits around the stable
Lagrangian points L4 and L5 \citep{1989A&ARv...1..261C} are stable in large
fractions of their characteristic in M10. Thus, the regions 
around L4,
L5 are populated and we observe the formation of banana-like features. These
regions do not reach the neighborhood of L1 and L2, because the p.o. of the
banana-like family at larger $E_j$ are unstable. As a result 
the regions close
to the unstable Lagrangian points are rather empty (unlike with what we observe
in Fig.~\ref{rm01hom} for model M1 and in Fig.~\ref{homom4} for model M4).
Despite the fact that banana-like orbits have been associated with the
reinforcement of spiral arms in some models \citep[see e.g. the model
of][for the Local Arm]{lep17},
such
morphological features are hardly observed in real galaxies, at least as strong
features. Two reasons that
could lead to the absence of these structures in real galaxies are (i) large
corotation distances from the center or (ii) strong perturbations. As we have
noted, Fig.~\ref{m10model}b is not weighted taking into account the exponential
character of the imposed disk and spirals. However, in images like
Fig.~\ref{m4_3dexpo1} the corotation region is a low density region and the
banana-like features are not discernible. The second option is that the families
around L4 and L5 are unstable. This can happen in \textit{barred}-spiral models,
where the forcing at the corotation region is strong. In the latter case,
particles located at the L4, L5 regions, are not trapped by regular orbits.
Their motion will be determined firstly by the presence of the stable branch of
the unstable manifold of the unstable p.o. around L1, L2. This 
will lead them to
the neighborhood of either L1 or L2 and then, the unstable branch of the 
manifold
will lead them beyond corotation, reinforcing ``chaotic'' spirals. This
mechanism for populating or depleting the areas around L4 and L5 in a response
model by changing the forcing has been presented in \citet{2017Ap&SS.362..129P}.
In Fig.~\ref{m10model}b we can observe weak ``chaotic'' spirals
emerging from the
L1 and L2 regions. The dynamical mechanisms that act in this case are similar to
those in the models in \citet{2017Ap&SS.362..129P}, where we have two sets of
spirals supported by regular (inside corotation) and chaotic orbits (beyond
corotation) coexisting (see e.g. their figure 2). 

\section{Conclusions}\label{sec:concl}
We have investigated the orbits that can support an open, thick (3D) spiral
structure in the PERLAS potential. We have taken advantage of the PERLAS
property to have everywhere positive density as the parameters of the spiral
potential part vary and we investigated models in the range $0.01  \leq M_s/M_d 
\leq 0.1$. Our main conclusions are the following:

\begin{enumerate}
\item A thick spiral pattern in the PERLAS potential is supported by orbits
associated with the stable vertical bifurcations of x1 at the vertical $n:1$
resonances with $n \geqq 3$. The thickness of the spirals supported by such
orbits in our models is about 0.6~kpc (i.e. 0.3~kpc above and 
below the
equatorial plane). Fig.~\ref{splot2} illustrates how the periodic orbits in our
model shape an appropriate backbone for reinforcing a thick spiral. Around
these p.o. can be build a 3D spiral.
\begin{figure}
\begin{center}
 \includegraphics[scale=0.5]{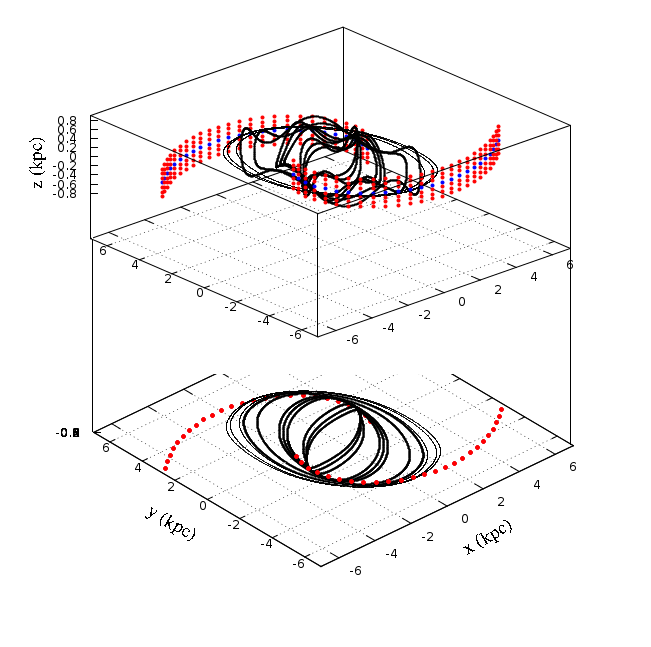}\\
\end{center}
    \caption{A set of spiral-supporting periodic orbits from model M4, which
reinforce a spiral pattern of thickness 0.6~kpc. In the upper part of the figure
we can see how the 3D orbits, by oscillating above and below the z=0 plane
reinforce a spiral ribbon, while in the lower panel we give the 
projections of the
p.o. and the spiral ribbon on the equatorial plane, $z=0$.}
    \label{splot2}
\end{figure}
\item By varying the mass ratio $M_s/M_d$ in our models we concluded that the
best relation between the imposed spiral and the orbital content of the models
is for $0.03 \lessapprox M_s/M_d \lessapprox 0.07$.
\item We find a thick, strong, bi-symmetric spiral pattern extending between the
ILR and the radial 4:1 resonance of the model in agreement with previous studies
by \citet{1986A&A...155...11C, 1988A&A...197...83C},
\citet{1991A&A...243..373P}, \citet{1994A&A...286...46P},
\citet{1996A&A...315..371P} and \citet{1997A&A...323..762P}. The characteristic
morphological feature associated with the 4:1 resonance in all models, was the
bifurcation of the arms.
\item Inside the ILR, although we do not have an explicit bar component in the
potential, the particles feel an m=2 term due to the existence of the massive
spirals starting, by construction, at the ILR. This leads to a bar-like
response. We find x1v1-like orbits supporting a 3D peanut structure in the
central region of the models. The backbone of the p.o. inside the ILR, projected
on the equatorial plane, gives always a continuation of the spirals in the
central part with an abrupt increase of their pitch angle.
\item Just after the radial 4:1 resonance we observe in general a gap at the
spiral region in the response models. However, between 4:1 and corotation we
find in the PERLAS potential local enhancements of the spiral arms, i.e.
segments of arms, close to the imposed spiral loci. Similar local enhancements
are not found beyond corotation. If we take into account the exponential profile
of the PERLAS model these segments of arms are found in sparsely populated
regions of the galaxies. Thus, it is unlikely to be part of the strong
symmetric arms observed in grand-design, normal spirals. Until now, extensions 
of the arms beyond
the 4:1 resonance were found in gaseous response models \citep[see e.g.
figure 4 in][]{1997A&A...323..762P}, while weak extensions have been identified
in $N$-body stellar models \citep[see e.g. figure 10 in][]{pk99}.
\item In our response models, especially in those with large $M_s/M_d$ ratios,
we observe ``chaotic'' spirals emerging from the L1, L2 regions. In such a case
in the models act the dynamical mechanisms described in
\citet{2017Ap&SS.362..129P} that support in the same model an inner and an outer
spiral structure. These outer spirals are in regions of low 
density in
the models.
\end{enumerate}
 L.C.V. and I.P. thank the Mexican Foundation CONACYT
for grants that supported this research. L.C.V. thanks the Research 
Center for Astronomy (RCAAM) of the Academy of
Athens for its hospitality during his visit there, when part of
this work has been completed. LCV thanks IA-UNAM for their hospitality
during his visit there. This work is part of the project
“Study of stellar orbits and gravitational potentials in galaxies, 
with numerical and observational methods” in which researchers from 
RCAAM and INAOE participate.



\end{document}